\documentclass[aps, prl, twocolumn, superscriptaddress, amsmath, tightenlines, longbibliography]{revtex4-1}

\usepackage{amssymb}
\usepackage{amsmath}
\usepackage{dcolumn}
\usepackage{graphicx}
\usepackage{mathrsfs}
\usepackage{appendix}
\usepackage{graphicx}
\usepackage{booktabs}
\usepackage{units}
\renewcommand{\Re}{\operatorname{Re}}

\def \hide#1{}

\newcommand{\ket}[1]{\mbox{$|#1\rangle$}}
\newcommand{\bra}[1]{\mbox{$\langle#1|$}}

\usepackage{url}
\usepackage[colorlinks]{hyperref}
\hypersetup{%
	plainpages=true,
	breaklinks=true,       
	hypertexnames=false,  
	pageanchor=true,
	colorlinks=true,
	linkcolor={blue},
	citecolor={red},
	urlcolor={blue},
	anchorcolor={black}
}

 \makeatletter

\newcommand{\Rmnum}[1]{\expandafter\@slowromancap\romannumeral #1@}
\makeatother

\usepackage{mleftright} 

\AtBeginDocument{%
	\newwrite\bibnotes
	\def\bibnotesext{Notes.bib}
	\immediate\openout\bibnotes=\jobname\bibnotesext
	\immediate\write\bibnotes{@CONTROL{REVTEX41Control}}
	\immediate\write\bibnotes{@CONTROL{%
			apsrev41Control,author="08",editor="1",pages="0",title="0",year="1"}}
	\if@filesw
	\immediate\write\@auxout{\string\citation{apsrev41Control}}%
	\fi
}%

\begin{document}

\title{Tunable Chiral Bound States with Giant Atoms}

\author{Xin Wang}
\affiliation{Theoretical Quantum Physics Laboratory, RIKEN Cluster for Pioneering Research, Wako-shi, Saitama 351-0198, Japan}
\affiliation{MOE Key Laboratory for Nonequilibrium Synthesis and Modulation of Condensed Matter, School of Physics, Xi'an Jiaotong University, 710049 Xi'an, P.R.China}

\author{Tao Liu}
\affiliation{Theoretical Quantum Physics Laboratory, RIKEN Cluster for Pioneering Research, Wako-shi, Saitama 351-0198, Japan}

\author{Anton Frisk Kockum}
\affiliation{Department of Microtechnology and Nanoscience, Chalmers University of Technology, 41296 Gothenburg, Sweden}

\author{Hong-Rong Li}
\affiliation{MOE Key Laboratory for Nonequilibrium Synthesis and Modulation of Condensed Matter, School of Physics, Xi'an Jiaotong University, 710049 Xi'an, P.R.China}

\author{Franco Nori}
\affiliation{Theoretical Quantum Physics Laboratory, RIKEN Cluster for Pioneering Research, Wako-shi, Saitama 351-0198, Japan}
\affiliation{Physics Department, The University of Michigan, Ann Arbor, Michigan 48109-1040, USA}
\date{\today}

\begin{abstract}

We propose tunable chiral bound states in a system composed of superconducting giant atoms and a Josephson photonic-crystal waveguide (PCW), with no analog in other quantum setups. The chiral bound states arise due to interference in the nonlocal coupling of a giant atom to multiple points of the waveguide. The chirality can be tuned by changing either the atom-waveguide coupling or the external bias of the PCW. Furthermore, the chiral bound states can induce directional dipole-dipole interactions between multiple giant atoms coupling to the same waveguide. Our proposal is ready to be implemented in experiments with superconducting circuits, where it can be used as a tunable toolbox to realize topological phase transitions and quantum simulations.

\end{abstract}

\maketitle


\paragraph*{Introduction.---}

Over the past decades, superconducting quantum circuits (SQCs) have emerged as a powerful platform for quantum information processing~\cite{You2003, Koch07,You051,rClarke08, Gu2017, Song2019, Arute2019, Kockum2019a, Kjaergaard2020}. For this development, the strong coupling that can be achieved between superconducting qubits (artificial atoms) and microwave photons has played an important role~\cite{Wallraff2004, Niemczyk2010, Kockum2019b}. Unlike conventional atom-light interaction, the atomic size in an SQC platform can be comparable to the wavelength of light, indicating that the dipole approximation is no longer valid~\cite{Gustafsson2014, Kockum2014, Aref2016, Guo2017, Kockum2018, Andersson2019, Kockum2019, Guo2019, Kannan2020}. Such atoms are called superconducting giant atoms. They are nonlocally coupled to multiple points of a waveguide~\cite{Kockum2014, Manenti2017, Delsing2019, Sletten2019, Kannan2020, Vadiraj2020, Andersson2020, Bienfait2020}. Interference effects between these points significantly modify the atom-matter interaction, and, therefore, change the collective behavior of the atoms~\cite{Kockum2014, Kockum2018, Kockum2019}. Furthermore, non-Markovian effects, due to time delay of waves propagating between distant coupling points, can play an important role in giant-atom dynamics~\cite{Guo2017, Andersson2019, Guo2019}. All these exotic phenomena have no counterpart in conventional atom-light systems.

Recently, a number of studies have explored chiral quantum phenomena in waveguide quantum electrodynamics, which enables cascaded quantum circuits, directional qubit interactions, and simulations of many-body physics~\cite{Carmichael1993, Gardiner93, Pichler2015, Ramos2016,Lodahl2017, Vermersch2017, Xiang2017, Grankin2018, Bello2019, Calaj2019,Bliokh2019, Guimond2020}. To achieve these chiral features, many approaches have been proposed for designing unidirectional waveguides, including subwavelength confinement in nanophotonic systems \cite{Mitsch2014,Bliokh2015, Petersen2014, Young2015, Bliokh2015a, Bliokh2015b}, spatiotemporal modulation~\cite{Calaj2019}, topological engineering~\cite{Bello2019, arXiv:2005.03802}, and structures integrated with synthetic gauge fields~\cite{Ramos2016, PhysRevResearch.2.023003}. The corresponding chiral quantum behavior can emerge via either real propagating photons or virtual nonradiative photons~\cite{Pichler2015, Ramos2016, Lodahl2017, Bello2019,Leonforte2020}. In particular, chiral quantum systems based on virtual photons can induce directional dipole-dipole interactions between qubits~\cite{Bello2019}, as demonstrated recently in an SQC experiment using a topological waveguide~\cite{arXiv:2005.03802}. Experimental realizations of most previous proposals remain elusive in SQCs, so the study of chiral quantum phenomena in SQCs is still in its infancy~\cite{Guimond2020}. Furthermore, the chiral interactions in previous proposals cannot be tuned well, which limits their applications in quantum information processing~\cite{Liao2010,Liu2017, Sundaresan19, Carusotto2020}. 

In this work, we present an alternative tunable chiral quantum system in SQCs. Its directional nature stems from interference effects due to nonlocal coupling of superconducting giant atoms to a Josephson photonic crystal waveguide (PCW). The PCW is constructed by a Josephson-chain metamaterial~\cite{Rakhmanov2008,Leib2012, Masluk2012, Altimiras2013, Rastelli2013, Weil2015, Krupko2018, Mirhosseini2018, weissl2014quantum, Karkar2019, Martinez2019, Planat2020}, which can be tuned via an external flux bias. The nonlocal coupling of a giant atom to two points of the PCW results in the appearance of chiral bound states, \emph{ whose chiralities can be freely tuned}. Tunable chiral many-body interactions of multiple giant atoms is realized through exchange of virtual photons between overlapping such bound states. 

\begin{figure}[tbph]
	\centering \includegraphics[width=8.5cm]{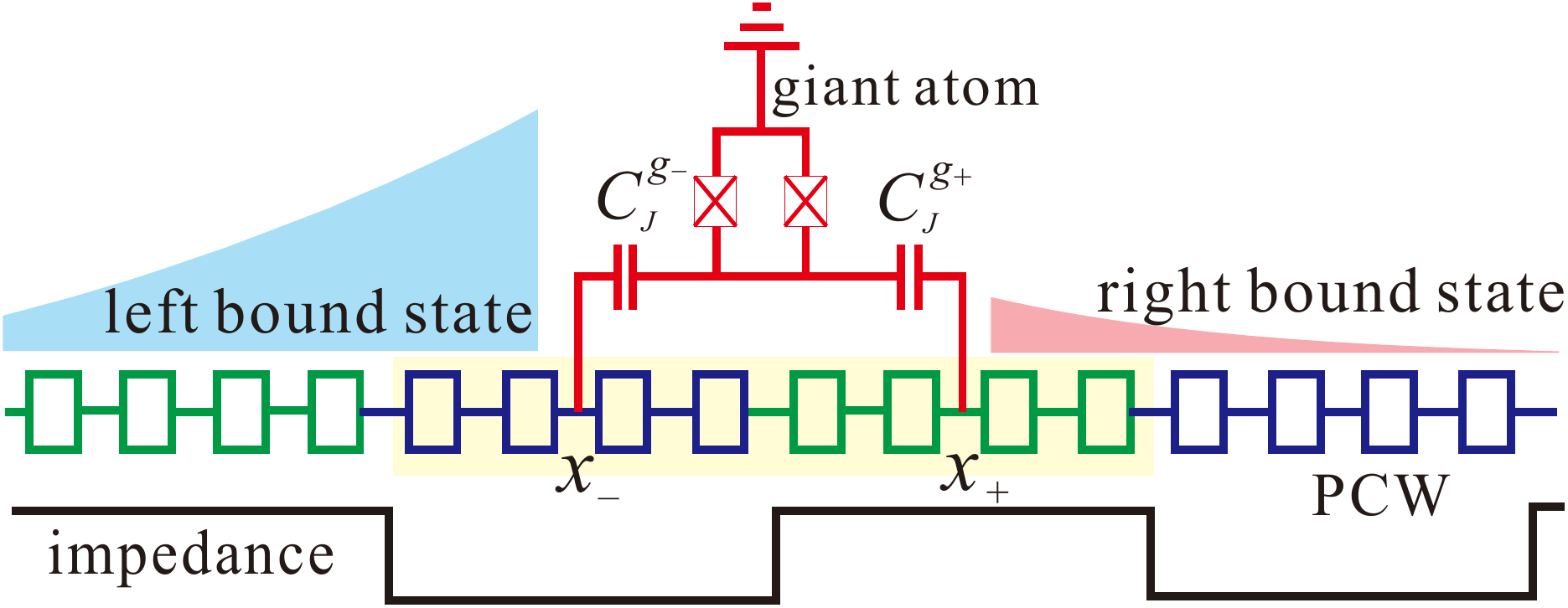}
	\caption{Sketch of the system. A superconducting giant atom (red) couples via capacitances $C_{J}^{g\pm}$ to two points $x_{\pm}$ of a PCW (squares, superconducting quantum interference devices [SQUIDs]). The green (blue) SQUIDs denote high (low) impedance. In our work, we assume the periodic impedance modulation as a cosine wave. The results remain robust when we consider other shapes of modulation signals (e.g., a square wave)~\cite{supplement}. The points $x_{+}$ and $x_{-}$ are assumed to be within one period of the modulation (yellow area). The left (right) photonic component of a bound state is shown in blue (red).}
	\label{fig1m}
\end{figure}

\paragraph*{Giant-atom-induced tunable chiral bound states.---}
As shown in Fig.~\ref{fig1m}, we consider a giant superconducting atom coupled to two points $x_{\pm}$ of a Josephson-chain PCW via capacitances $C_{J}^{g\pm}$. In contrast to conventional nanophotonic waveguides~\cite{John1990,Zhou2008, John1991, Hung2013, Goban2014, GonzlezTudela2015, Douglas2015, Hood2016, Douglas2016, Munro2017, Chang2018}, this microwave PCW has a wide range of tunable parameters, including  the unit cell length and impedance controlled by the external flux \cite{Johansson09L, Johansson2010B, Pogorzalek17, Wang2019}. The detailed construction and spectrum of a Josephson PCW can be found in Secs.~$\textrm{\Rmnum{1}}$ and $\textrm{\Rmnum{2}}$ in the Supplementary Material~\cite{supplement}.

When the atomic transition frequency $\omega_q$ is in the PCW bandgap, and close to the top of the lowest energy band $\omega_k$ in the first Brillouin zone (BZ), the interaction Hamiltonian is
\begin{equation}
H_{\text{int}} = \sum_{k \in \text{BZ}} \hbar \Delta_k a^\dag_k a_k 
+ \sum_{k \in \text{BZ}} \hbar \mleft( g_k a^\dag_k \sigma_- + g^*_k a_k \sigma_+ \mright),
\label{Hrtot}
\end{equation}
where $\Delta_k = \omega_k - \omega_q$ is the frequency detuning,  $\sigma_{\pm}$ are the atom raising and lowering operators, $a_k$ ($a_k^\dag$) is the annihilation (creation) operator of the photonic mode with wavevector $k$ in the lowest energy band, and the nonlocal atom-waveguide coupling strength is given by (see Sec.~$\textrm{\Rmnum{3}}$ in Ref.~\cite{supplement})
\begin{gather}
g_k = \sum_{i = \pm} g_k^i e^{i k x_i} u_k (x_i), ~~\textrm{with}~~  g_k^\pm \simeq \frac{e}{\hbar} \frac{C_J^{g \pm}}{C_\Sigma}\sqrt{\frac{\hbar \omega_q}{C_t}}.
\label{gkG}
\end{gather}
Here $C_{\Sigma}$ ($C_t$) is the total capacitance of the atom (PCW), and $u_k (x) = u_k (x + \lambda_m)$ is the Bloch wavefunction with a tunable modulating wavelength $\lambda_m$ of the PCW. We compute $u_k (x)$ and $g_k$ numerically based on experimental values of the SQUID-chain PCWs in Refs.~\cite{weissl2014quantum, Krupko2018, Martinez2019}.

We find the bound state of the system by solving $H_{\text{int}}\ket{\psi_b} =\hbar \epsilon_b \ket{\psi_b}$, with eigenenergy $\epsilon_b$ and eigenstate $\ket{\psi_b} = \cos(\theta) \ket{e,0} + \sin(\theta) \sum_k c_k  \ket{g,1_k}$, in the single-excitation subspace. 
Previous studies of small atoms in waveguides~\cite{Hung2013, Goban2014, GonzlezTudela2015, Douglas2015, Hood2016, Douglas2016, Munro2017}, and one on giant atoms in an coupled-resonator waveguide~\cite{Zhao2020}, have shown that the bound state decays exponentially and symmetrically in both directions.  

For the case of a giant atom, the real-space wavefunction of the photonic component of the bound state is approximated by (see Sec.~$\textrm{\Rmnum{4}}$ in \cite{supplement})
\begin{equation}
\phi_b (x) = \sin (\theta) \bra{x} \sum_k c_k a_k^\dag \ket{0} = \sum_{i = \pm} \phi_b^i (x),
\label{phi_inte}
\end{equation}
with
\begin{equation}
\phi^i_b (x) \varpropto \!\!\int\!\! \frac{g_k^i u_k (x_i) u_k^* (x) e^{-i k(x-x_i)}}{\epsilon_b - \Delta_k} dk \!=\! A^i (x) e^{i \theta_i (x)},
\label{phibound0}
\end{equation}
where $\phi^\pm_b (x)$ represent the photonic wavefunction components of the bound state for a small atom coupling at positions $x_\pm$; $A^\pm$ and $\theta_\pm$ denote their corresponding amplitudes and phases, respectively. The phases of atom-waveguide coupling amplitudes [see Eq.~(\ref{gkG})] at positions $x_+$ and $x_-$ cannot be simultaneously gauged out due to the nonlocal coupling. This results in interference between $\phi^+_b (x)$ and  $\phi^-_b (x)$ [see Eqs.~(\ref{phi_inte}), (\ref{phibound0})], leading to the formation of a chiral bound state.

\begin{figure}[tb!]
	\centering \includegraphics[width=8.6cm]{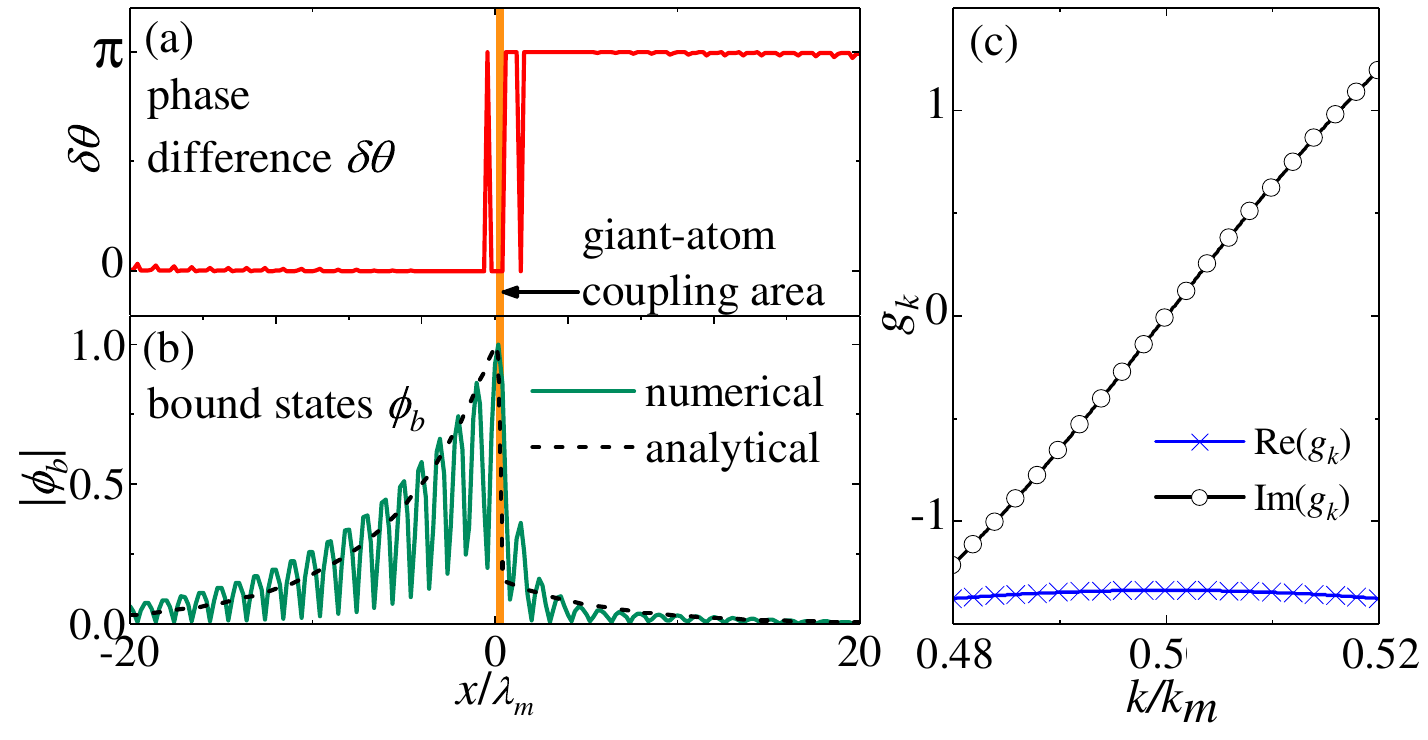}
	\caption{Properties of a chiral bound state. (a) Phase differences between $\phi^{+}_{b}(x)$ and  $\phi^{-}_{b}(x)$ versus of $x$ for $\{x_-, x_+\} = \{0, 0.5\lambda_{m}\}$. (b) The bound state amplitude $|\phi_b(x)|$ for the same setup. The solid (dashed) curve is the numerical (analytical) result described by Eq.~(\ref{phibound0})  [Eq.~(\ref{phib_ana})]. (c) The imaginary and real part of $g_k$ versus $k$ for the same setup.}
	\label{fig2m}
\end{figure}

As depicted in Fig.~\ref{fig1m},  we assume the PCW to be infinitely long in both directions, and the original point $x=0$ at the middle of the low-impedance part in one cell.
The giant-atom coupling points $x_-$ and $x_+$ are located at $\{x_-, x_+\} = \{0, 0.5\lambda_{m}\}$.
We find that the bound-state components $\phi^{\pm}_b(x)$ distribute symmetrically around the coupling points~\cite{supplement}. However, as depicted in Fig.~\ref{fig2m}(a), their phase difference is $\delta\theta = \theta_+ - \theta_- \simeq 0$ for $x\ll x_-$, while it is $\delta\theta \simeq \pi$ for $x\gg x_+$.
Numerical results indicate that, maximum interference is achieved when $g_k^+ \simeq 3.4 g_k^-$, giving $A^+(x) \simeq A^-(x)$.
Consequently, the right (left) bound state vanishes (is maximally enhanced) due to destructive (constructive) interference. Figure \ref{fig2m}(b) shows the real-space distribution of $\phi_{b}(x)$, which is strongly localized to the left of the giant atom. 
Note that, for other positions of the coupling points, there exist different interference patters between $\phi^{\pm}_{b}(x)$, which lead to different chirality. Detailed discussions can be found in Sec.~$\textrm{\Rmnum{4}}$ in Ref.~\cite{supplement}.

The chiral bound state can be phenomenologically interpreted as a result of interference, as explained above. We now make a quantitative analysis. When a small atom is coupled to a PCW, as studied in Refs.~\cite{Douglas2015, Hood2016, Douglas2016, Munro2017, Chang2018}, the atom-waveguide coupling amplitude $g_k$ is a constant independent of the wavevector $k$, i.e., $g_k \simeq g_{k_0}$, with $k_0 = k_m/2$, and $k_m = 2\pi / \lambda_m$. The giant-atom case is different, as shown in Fig.~\ref{fig2m}(c), where we plot the real and imaginary parts of $g_k$ versus $k$. Note that, around $k_0$, the real part of $g_k$ is approximately constant, but the imaginary part changes linearly with $k$. We therefore rewrite $g_k$ as
\begin{equation}
g_{k}\simeq (A + i B \delta k),
\label{gkri}
\end{equation}
where $\delta k = k - k_0$, $A$ represents the average real part of $g_k$ around $k_0$, and $B$ is the slope of the imaginary part of $g_k$. Due to the \emph{nonlocal coupling} of the giant atom to the PCW, $B$ has a non-zero value, and cannot be gauged out. In addition, by considering the effective-mass approximation~\cite{Douglas2015, GonzlezTudela2015}, the dispersion relation of the lowest energy band of the PCW around the band edge can be expressed as $\Delta_k = -\delta_0 - \alpha_m (k-k_0)^2$ (see Sec.~$\textrm{\Rmnum{3}}$ in Ref.~\cite{supplement}), and $\phi_b (x)$ in Eq.~(\ref{phi_inte}) becomes
\begin{equation}
\phi_b (x) \varpropto \mleft[ C_- \Theta(-x) + C_+ \Theta(x) \mright] \exp \mleft( -\frac{|x|}{L_{\text{eff}}} \mright),
\label{phib_ana}
\end{equation}
where $L_{\text{eff}}=\sqrt{\alpha_{m}/\delta_{0}}$ is the decay length, $\Theta(x)$ is the Heaviside step function, and $C_{\pm}$ are determined by the imaginary and real parts of $g_{k}$ as
\begin{equation}
C_{\pm} = A \pm B \sqrt{\frac{\delta_0}{\alpha_m}}.
\label {cpm}
\end{equation}

In Eq.~(\ref{phib_ana}), we have assumed $|x_+ - x_-| < \lambda_m \ll L_{\text{eff}} $. This approximation holds in Fig.~\ref{fig2m}(b), where the photonic component between the two coupling points (brown area) is much smaller than the left and right parts, and can be neglected. Therefore, when considering the bound-state distribution, we view $x_{\pm}$ as both approximately being at $x = 0$. For the parameters used in Fig.~\ref{fig2m}, we have $|C_-| \gg |C_+|$. Consequently, the photonic component of the bound state mostly distributes to the left of the giant atom. Note that the above analytical results fit well with the numerical ones, as shown in Fig.~\ref{fig2m}(b).

We now define the chirality of the bound state as
\begin{equation}
\mathcal{C}_b = \frac{\Phi_L - \Phi_R}{\Phi_L + \Phi_R}, \quad \Phi_{R/L} = \mleft| \int_{\pm \infty}^{x_{\pm}} |\phi_b (x')|^2 dx' \mright|,
\label{chiral_def}
\end{equation}
where the chiral preferred direction is left (right) given that $\mathcal{C}_{b}>0$ ($\mathcal{C}_{b}<0$), and $\mathcal{C}_{b}\rightarrow 1$ ($\mathcal{C}_{b}\rightarrow -1$) indicates perfect left (right) chirality. Using Eq.~(\ref{phib_ana}), the analytical form of $\mathcal{C}_b$ becomes
\begin{equation}
\mathcal{C}_b = \frac{C_-^2 - C_+^2}{C_-^2 + C_+^2}.
\label{chiral_def2}
\end{equation}
In Fig.~\ref{fig3m}(a), we plot both the numerical and analytical $\mathcal{C}_b$ versus $x_+$ by fixing $x_- = 0$ and $g_k^+ = g_k^-$. The numerical parameters of the whole system are adopted from the experiments in Refs.~\cite{weissl2014quantum, Krupko2018, Martinez2019} (see Sec.~$\textrm{\Rmnum{3}}$ in Ref.~\cite{supplement}).
The sign of {$\mathcal{C}_{b}$ changes when the second coupling point $x_+$ is moved from the left side of $x_-$ to the right side. At $x_+ / \lambda_m \simeq \pm 0.7$ [dashed lines in Fig.~\ref{fig3m}(a)], the chirality reaches its maximum value $|\mathcal{C}_b| \simeq 0.95$.
	In Fig.~\ref{fig3m}(b), we fix $\{ x_-, x_+ \} = \{ 0, 0.75\lambda_m \}$ and plot $\mathcal{C}_{b}$ versus the coupling strength $g_k^+$. The results in Fig.~\ref{fig3m} show that \emph{the chirality can be continuously tuned} over the whole range $\mathcal{C}_{b}\in (-1,1)$ by changing either $x_+$ or $g_k^+$.
	
	\begin{figure}[tb!]
		\centering \includegraphics[width=8.6cm]{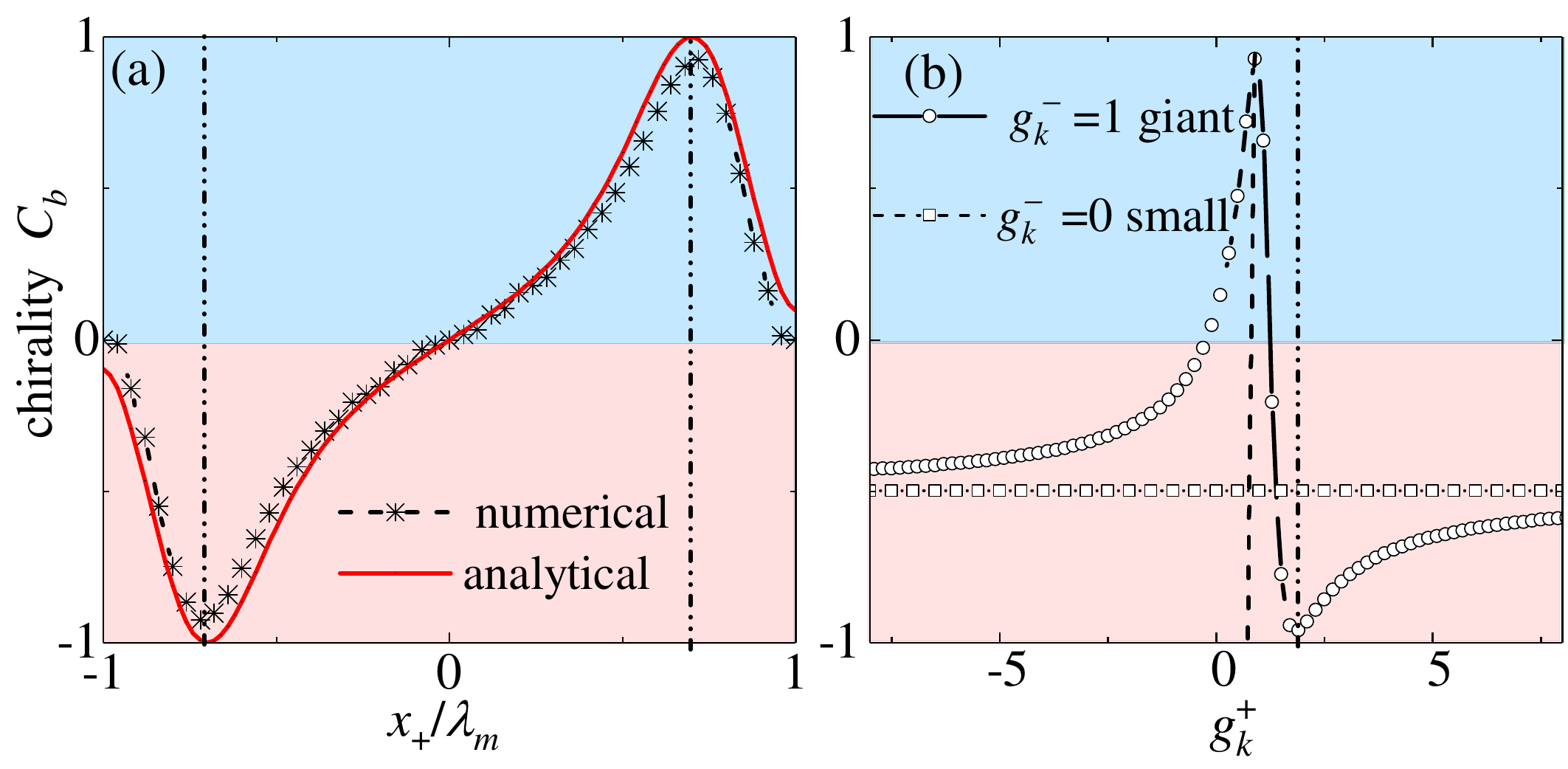}
		\caption{Tuning the chirality of the bound state. (a) Chirality $\mathcal{C}_b$ versus $x_+$ for fixed $x_- = 0$ and $g_k^+ = g_k^-$. The dashed lines indicate the maximum chirality $|\mathcal{C}_b| \simeq 1$. (b) Chirality versus $g_k^+$ for fixed $\{ x_-, x_+ \} = \{0, 0.75\lambda_m \}$ and $g_k^- = 1$ (giant-atom case) or $g_k^- = 0$ (small-atom case). }
		\label{fig3m}
	\end{figure}
	
	Unlike the case of nanophotonic waveguide quantum electrodynamics~\cite{Hung2013, Chang2018}, an atom placed in the PCW can see different semi-infinite waveguide structures to the right and left of a coupling point. This symmetry-breaking is what enables chirality. As shown in Fig.~\ref{fig3m}(b), the symmetry-breaking enables the formation of a chiral bound state not only with a giant atom, but also with a small atom, which has not been explored in previous studies.
	For $|g_k^+| \gg |g_k^-|$, which corresponds to the smal-atom case, we find $\mathcal{C}_{b} \simeq -0.49$ when $\{ x_-, x_+ \} = \{0, 0.75\lambda_m \}$. However, the chirality for a small atom is never perfect ($|\mathcal{C}_b| < 1$) and cannot be tuned by changing the coupling strength, in contrast to the giant-atom case [Fig.~\ref{fig3m}(b)]. Additionally, $\mathcal{C}_{b}$ is quite sensitive to the coupling position. A detailed discussion is provided in Ref.~\cite{supplement}.
	
	
	\paragraph*{Chiral dipole-dipole interactions.---}
	
	When multiple small atoms are coupled within the bandgap of the same PCW, effective dipole-dipole interactions between them are induced through the exchange of virtual photons in the PCW. The interaction strength is determined by the overlap between the decaying evanescent fields of the bound states~\cite{Goban2014, GonzlezTudela2015, Douglas2015, Hood2016, Douglas2016, Munro2017, Chang2018}. In previous studies with cold-atom systems~\cite{Douglas2015, Chang2018}, where the atoms are equally spaced, the nearest-neighbor interaction strength is constant. For the chiral bound states induced by the coupling between a PCW and giant atoms, the scenario is different.
	
	As shown in Fig.~\ref{fig4m}, we consider giant atoms A and B equally distributed along a PCW with an inter-atom distance $D_q$. Here, the PCW impedance modulation is simplified as square waves. One leg of each giant atom A and B is coupled to the low-impedance points $x^{A,B}_-$. while the second coupling points $x_+^{A,B}$ are placed either to the left or to the right (different for A and B) of $x^{A,B}_{-}$ at the closest high-impedance position. Therefore, the bound states of giant atoms A and B satisfy $\mathcal{C}^A_b = -\mathcal{C}^B_b$.
	The atomic pair $\{A, B\}$ is repeated along the PCW, and can be viewed as a dimer (see Fig.~\ref{fig4m}). The intracell dipole-dipole interaction can be derived via standard resolvent-operator techniques~\cite{cohen1998atom, GonzlezTudela2017, supplement}:
	\begin{equation}
	\text{Re}[\Sigma_{AB} (z)] = \text{Re}\mleft[\int_0^{k_0} dk \frac{2 \Re(g_{kA} g_{kB}^*)}{z - \Delta_k}\mright] \simeq J_{AB},
	\label{sig12}
	\end{equation}
	where the real part $J_{AB}$ of $\Sigma_{AB} (z)$ describes the coherent dipole-dipole coupling between intracell atoms. The atomic decay effects can be strongly suppressed with a large detuning $\delta_{0}$~~\cite{GonzlezTudela2017,Bello2019,Ramos2016} (see Sec.~V in Ref.~\cite{supplement}). The intercell coupling amplitude $J_{BA}$ can be found similarly.
	
	\begin{figure}[tb]
		\centering \includegraphics[width=8.6cm]{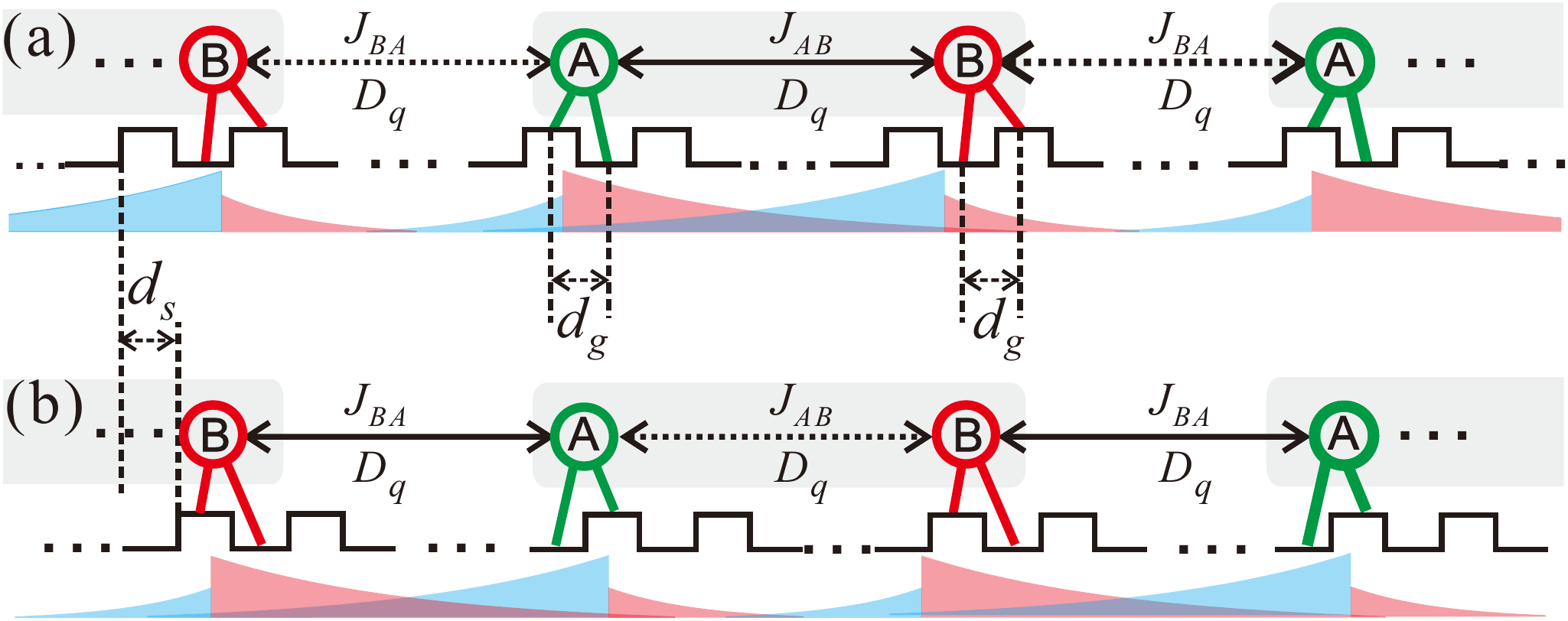}
		\caption{Setups for chiral dipole-dipole interactions betweeen giant atoms. (a) Giant atoms A and B, separated by a distance $D_q$, coupled to a Josephson-chain PCW. The periodic impedance modulation is a square wave. The induced effective dipole-dipole coupling is chiral with $J_{AB} \gg J_{BA}$. (b) Relative to (a), the impedance modulation signal is shifted with a distance $d_s = 0.5 \lambda_m$. The bound-state chiralities are reversed, resulting in $J_{AB} \ll J_{BA}$.}
		\label{fig4m}
	\end{figure}
	
	In Fig.~\ref{fig5m}(a), we numerically plot $J_{AB}$ and $J_{BA}$, both of which exponentially decay with $D_q$~\cite{Goban2014, GonzlezTudela2015, Douglas2015, Hood2016, Douglas2016}, for the setup in Fig.~\ref{fig4m}(a). Since the bound-state chiralities of A and B are opposite, the decaying evanescent fields within a unit cell have much larger overlap than those between different cells. This leads to a much larger intracell dipole-dipole interaction $J_{AB}$ than the intercell interaction $J_{BA}$, i.e., the interaction is \emph{chiral} even though the giant atoms are equally spaced. Since the bound states of A and B can be tuned to $|\mathcal{C}_b|\simeq1$, the atoms only interact with the atoms in their chiral preferred directions, but cannot interact with those in the opposite direction, no matter how small the separation $D_q$ is.
	
	\begin{figure}[tb]
		\centering \includegraphics[width=8.6cm]{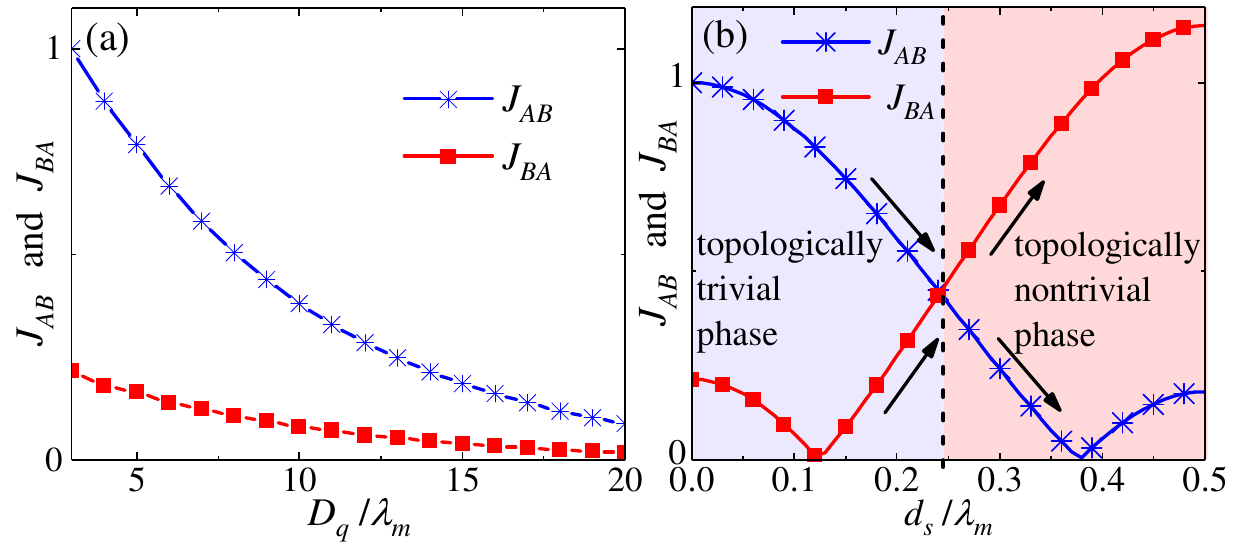}
		\caption{Chiral interaction and topological phase transition. (a) Dipole-dipole interaction strengths $J_{AB}$ and $J_{BA}$ (normalized by $J_{AB}$ at $D_q = 3\lambda_{m}$) versus separation $D_q$.
			(b) $J_{AB}$ and $J_{BA}$ for fixed $D_q$ versus of the shift distance $d_{s}$ [relative to Fig.~\ref{fig4m}(a)] for the PCW impedance modulation. As the shift distance is increased towards $d_s = 0.5 \lambda_m$, the chiral interaction strengths of $J_{AB}$ and $J_{BA}$ are exchanged, leading to a topological phase transition.}
		\label{fig5m}
	\end{figure}
	
	
	\paragraph*{Topological phases with giant atoms.---}
	
	The impedance of the Josephson PCW can be modulated via an external flux bias~\cite{supplement}. As shown in Fig.~\ref{fig4m}, shifting the programmable modulation signal by $d_s$, the high-impedance positions will be moved~\cite{supplement}. For $d_s  = 0.5 \lambda_m$, the chiralities of the giant atoms in Fig.~\ref{fig4m} are switched, leading to $J_{AB} \ll J_{BA}$. Figure.~\ref{fig5m}(b) shows that $J_{AB}$ ($J_{BA}$) decreases (increases) linearly with $d_s$ around $d_s \simeq 0.25\lambda_m$.
	
	By modulating $d_s$ in time, we can simulate topological phases. We assume that the frequency of each atom is also modulated in time~\cite{Gu2017}. As shown in Fig.~\ref{fig4m}, after tracing out the PCW, we map the atomic-chain Hamiltonian to the Su-Schrieffer-Heeger (SSH) model
	\begin{eqnarray}
	H_{\text{qc}} &=& \sum_i \mleft[ J_{AB}(t) \sigma_{Ai}^- \sigma_{Bi}^+ + J_{BA}(t) \sigma_{Bi}^- \sigma_{Ai+1}^+ + \text{h.c.} \mright] \notag \\
	&&+ \sum_i \Delta_q(t) (\sigma_{Ai}^z - \sigma_{Bi}^z), 
	\label{RM}
	\end{eqnarray}
	where $\Delta_q(t)$ is the frequency detuning between atoms A and B. The degeneracy point of the chain is at $\{J_{BA} - J_{AB}, \Delta_{q}\} = \{0, 0\}$~\cite{supplement}. The adiabatic Thouless pump trajectories (see Fig.~\ref{fig6m}), which encircle the degeneracy point, are topologically equivalent, and robust to disorders and perturbations~\cite{Lohse2015, Gu2017arxiv, Nakajima2016}. 
	As shown in Fig.~\ref{fig5m}(b), for $d_s \simeq 0.25 \lambda_m$, the bound states of atom A and B, do not show any chirality, leading to $J_{AB} - J_{BA} = 0$. This corresponds to the topological phase transition point~\cite{Su1979, Xiao2014, SaeiGharehNaz2018}.
	
	In Fig.~\ref{fig6m}(b), we plot the adiabatic pumping process for the evolution of an initial excitation localized at the left edge of the atomic chain (see Sec.~$\textrm{\Rmnum{5}}$ in Ref.~\cite{supplement}). At the end of each pump circle, the excitation is transferred to the right edge state with a high fidelity due to topological protection~\cite{Lohse2015}. This process exploits the directional interactions between giant atoms, and just needs to shift the modulation signal by a small length, which is feasible in SQC platforms.
	
	\begin{figure}[tb]
		\centering \includegraphics[width=8.6cm]{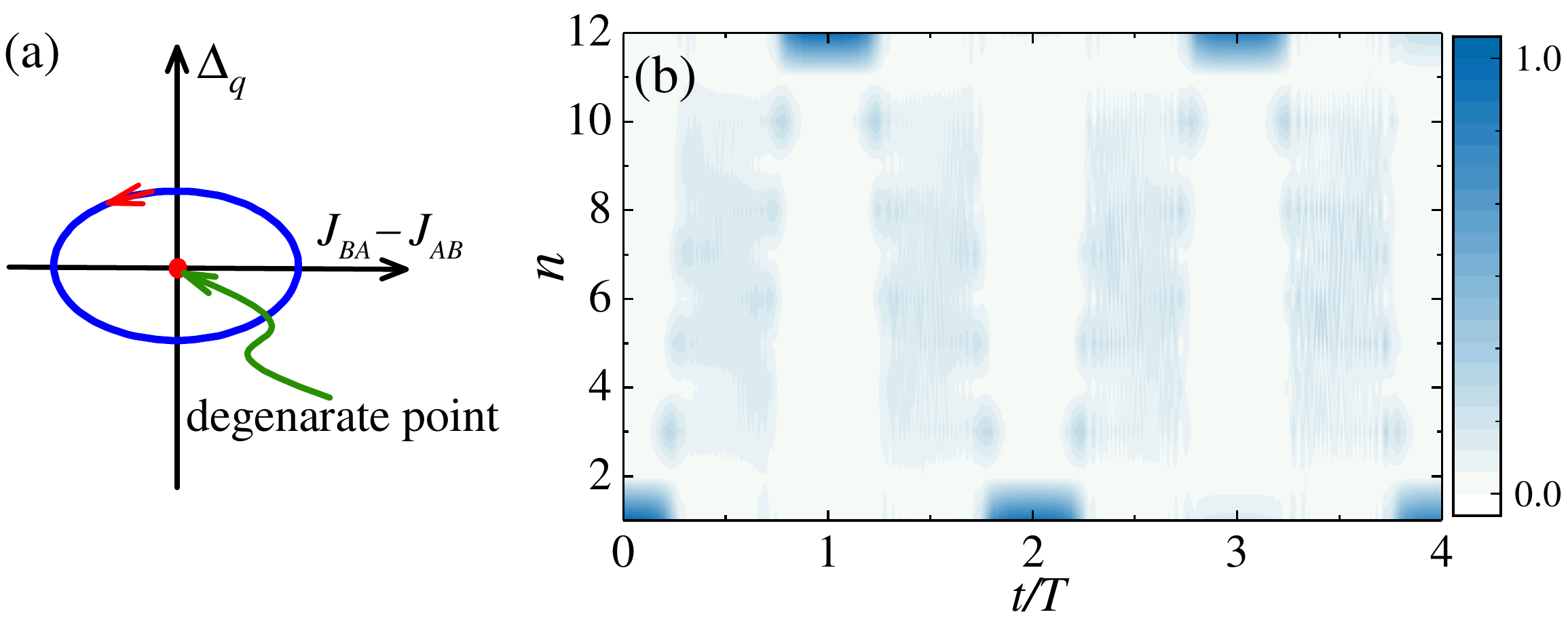}
		\caption{Topological protection. (a) The pump circle in parameter space  $\{J_{BA} - J_{AB}, \Delta_{q}\}$. The topologically nontrivial pumping corresponds to a closed path encircling the degeneracy point at the origin. (b) Time evolution of a single excitation in the atomic chain (with 12 sites) under adiabatic pumping loops (see Sec.~$\textrm{\Rmnum{5}}$ in Ref.~\cite{supplement}).}
		\label{fig6m}
	\end{figure}
	
	
	\paragraph*{Conclusion.---}
	
	In this work, we have explored giant superconducting atoms coupled to two points of a Josephson-chain PCW.
	We showed that interference arising due to the nonlocal coupling leads to chiral bound states. The chirality of these states can be easily tuned over the full range by either modulating the external flux bias of the PCW or changing the coupling strengths.
	For multiple giant atoms equally spaced along the waveguide, the dipole-dipole interactions exhibit strong chirality due to spatially asymmetric overlaps between the bound states. Each atom can be tuned to only interact with atoms in a preferred direction.
	Using this chiral interaction, we demonstrated that our proposal can realize a topological phase transition and topological Thouless pumping.
	Extending our setup to 2D PCWs might lead to more exotic quantum phenomena.
	We hope that our proposal can be a powerful toolbox to achieve chiral long-range interactions for quantum simulations and many-body physics. The setups we have studied here can be realized in experiments using currently available state-of-the-art technology for superconducting circuits.
	
		\emph{Acknowledgments.---}
		X.W.~is supported by
		China Postdoctoral Science Foundation No.~2018M631136
		and 
		the Natural Science Foundation of China under Grant No.~11804270.
		A.F.K.~acknowledges support from
		the Japan Society for the Promotion of Science (BRIDGE Fellowship BR190501),
		the Swedish Research Council (Grant No.~2019-03696),
		and
		the Knut and Alice Wallenberg Foundation through the Wallenberg Centre for Quantum Technology (WACQT).
		F.N. is supported in part by: NTT Research,
		Army Research Office (ARO) (Grant No. W911NF-18-1-0358),
		Japan Science and Technology Agency (JST)
		(via the Q-LEAP program and CREST Grant No. JPMJCR1676),
		Japan Society for the Promotion of Science (JSPS) (via the KAKENHI Grant No. JP20H00134
		and the JSPS-RFBR Grant No. JPJSBP120194828),
		the Asian Office of Aerospace Research and Development (AOARD),
		and the Foundational Questions Institute Fund (FQXi) via Grant No. FQXi-IAF19-06.

\let\oldaddcontentsline\addcontentsline
\renewcommand{\addcontentsline}[3]{}

\let\addcontentsline\oldaddcontentsline

\onecolumngrid

\newcommand\specialsectioning{\setcounter{secnumdepth}{-2}}
\setcounter{equation}{0} \setcounter{figure}{0}

\setcounter{table}{0} 
\renewcommand{\theequation}{S\arabic{equation}}
\renewcommand{\thefigure}{S\arabic{figure}}
\renewcommand{\bibnumfmt}[1]{[S#1]}
\renewcommand{\citenumfont}[1]{S#1}
\renewcommand\thesection{S\arabic{section}}
\renewcommand{\baselinestretch}{1.2}

\renewcommand{\theequation}{S\arabic{equation}}

\newpage

\setcounter{page}{1}\setcounter{secnumdepth}{3} \makeatletter
\begin{center}
{\Large \textbf{ Supplementary Material for\\
		Tunable Chiral Bound States with Giant Atoms}}
\end{center}

\begin{center}
Xin Wang$^{1,2}$, Tao Liu$^{1}$, Anton Frisk Kockum$^{3}$, Hong-Rong Li$^{2}$ and Franco Nori$^{1,4}$
\end{center}

\begin{minipage}[]{16cm}
\small{\it
	\centering $^{1}$Theoretical Quantum Physics Laboratory, RIKEN Cluster for Pioneering Research, Wako-shi, Saitama 351-0198, Japan \\
	\centering $^{2}$MOE Key Laboratory for Nonequilibrium Synthesis and Modulation of Condensed Matter, School of Physics, Xi'an Jiaotong University, 710049, P.R.China\\
	\centering $^{3}$Department of Microtechnology and Nanoscience, Chalmers University of Technology, 41296  Gothenburg, Sweden\\
	\centering $^{4}$Physics Department, The University of Michigan, Ann Arbor, Michigan 48109-1040, USA \\}
\end{minipage}

\vspace{8mm}

\begin{quote}

This supplementary material includes the following:
In Sec.~I, we discuss the lumped-circuit model of a Josephson chain working as a metamaterial SQUID transmission line (STL), and find the parameter regime where the linear dispersion relation is valid. In Sec.~II, we show how to realize a tunable photonic crystal waveguide (PCW) by periodic modulation of the STL's Josephson inductance via an external flux bias. In Sec.~III, we discuss the coupling between the PCW and a superconducting giant atom, and derive the analytical form for the chiral bound states. In Sec.~IV, we discuss the chiral bound state resulting from the interference effect due to nonlocal coupling of the giant atom. In Sec.~V we derive, by employing standard resolvent-operator techniques, the chiral dipole-dipole interactions between giant atoms mediated by virtual photons, and discuss how to realize topological pumping of the atomic chain by shifting the modulating signal of the PCW.

\end{quote}

\tableofcontents


\section{Dispersion relation of a SQUID transmission line}

\begin{figure}[tbph]
	\centering \includegraphics[width=\linewidth]{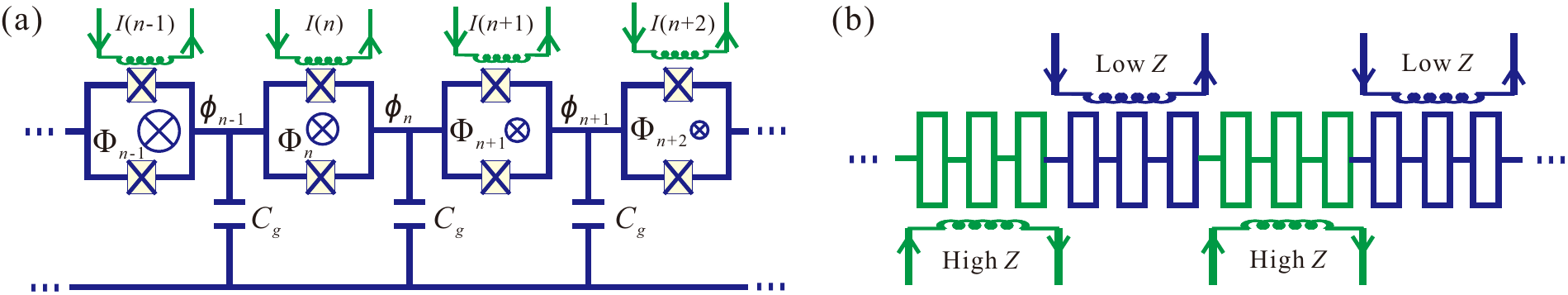}
	\caption{(a) A SQUID-chain platform for waveguide quantum electrodynamics involving superconducting giant atoms: each SQUID works as a tunable inductance which is controlled by a bias current $I(n)$ produced by an external coil. The Josephson capacitance $C_j$ (not shown) is in parallel with the Josephson inductance, and can be neglected in the linear-dispersion regime. At position $n$, the node flux is denoted by $\phi_n$, with a capacitance $C_g$ connecting to ground. (b) Instead of modulating the inductance site-by-site, the inductance of the SQUID chain (represented here by rectangles) can also be modulated group-wise. The periodic high and low impedances $Z$ can be tuned by a common current coil, which will produce a photonic-crystal waveguide (PCW) structure.
	}
	\label{fig1s}
\end{figure}

As shown in Fig.~\ref{fig1s}(a), we consider a microwave transmission line composed of a chain of $N$ superconducting quantum interference devices (SQUIDs), with capacitances $C_g$ connecting each node to the ground. The SQUIDs are separated by an equal spacing $d_0$. The $j$th SQUID can be viewed as a lumped inductance $L_j$, in parallel with the Josephson capacitance $C_j$~\cite{Masluk2012s, Altimiras2013s, weissl2014quantums, Weil2015s, Mirhosseini2018s, Karkar2019s, Martinez2019s}. The relation between $L_j$ and the external flux $\Phi_j$ is~\cite{Johansson09Ls, Pogorzalek17s, Wang2019s}
\begin{equation}
L_j = \frac{L_0}{\cos\mleft|\frac{\pi \Phi_j}{\Phi_0} \mright|}, \quad L_0 = \frac{\Phi_0^2}{8 \pi^2 E_{s0}}, 
\label{Lj0}
\end{equation}
where $E_{s0}$ is the junction Josephson energy, which is assumed to be identical for each cite, and $\Phi_0$ is the flux quantum. Alternatively, as indicated in Fig.~\ref{fig1s}(b), a group of SQUIDs can be tuned by sharing the same current coil. Denoting the flux at node $j$ as $\phi_j$, we obtain a Kirchoff current equation for the SQUID chain:
\begin{equation}
C_g \ddot{\phi}_j + \frac{\phi _j - \phi _{j-1}}{L_j} + C_j \mleft( \ddot{\phi}_j - \ddot{\phi}_{j-1} \mright) - \frac{\phi_{j+1} - \phi_j}{L_{j+1}} - C_j \mleft( \ddot{\phi}_{j+1} - \ddot{\phi}_j \mright) = 0.
\label{diff1}
\end{equation}
By assuming the capacitances and effective inductances identical, $C_j = C_J$ and $L_j = L_J$, the dynamical equation in Eq.~(\ref{diff1}) leads to the Hamiltonian~\cite{Weil2015s, Mirhosseini2018s}
\begin{eqnarray}
H_0 = \frac{1}{2} \vec{Q}^T \widehat{C}^{-1} \vec{Q} + \frac{1}{2} \vec{\Phi }^T \widehat{L}^{-1} \vec{\Phi},\\
\vec{\Phi }^T = \left(\phi_0, \phi_1, \ldots, \phi_N \right), 
\quad \vec{Q} = \hat{C} \dot{\vec{\Phi}},
\label{H0}
\end{eqnarray}
where the capacitance and inductance matrices are given by
\begin{eqnarray}
\widehat{C}&=&
\begin{pmatrix}
C_J & -C_J & 0 & \ldots &	&	\\
-C_J & 2C_J + C_g & -C_J & 0 &  \ldots &	\\
0 & -C_J & 2C_J + C_g & -C_J & 0 & \ldots \\ 
\vdots & 0 & \ddots & \ddots & \ddots & \ddots
\end{pmatrix},
\end{eqnarray}
and
\begin{eqnarray}
\widehat{L}^{-1}&=&
\begin{pmatrix}
\frac{1}{L_{J}} & -\frac{1}{L_{J}} & 0 & ... &	&	\\
-\frac{1}{L_{J}} & \frac{2}{L_{J}} & -\frac{1}{L_{J}} & 0 &  ... &	\\
0 & -\frac{1}{L_{J}} & \frac{2}{L_{J}} & -\frac{1}{L_{J}} & 0& ... \\ 
\vdots & 0 & \ddots & \ddots & \ddots &	\ddots
\end{pmatrix}.
\end{eqnarray}
Using the transformation $\vec{\psi}^s_k = \widehat{C}^{1/2} \vec{\Phi}$, the eigenfrequency $\omega_k$ for the system can be derived from~\cite{Weil2015s}
\begin{equation}
\widehat{C}^{-1/2} \widehat{L}^{-1} \widehat{C}^{-1/2} \vec{\psi}_k^s = \omega_k^2 \vec{\psi}_k^s,
\label{eigenp}
\end{equation}
where $\vec{\psi}_k^s$ is the wavefunction for mode $k$ with frequency $\omega_k$. 

As derived in Ref.~\cite{Weil2015s}, by assuming an open-ended boundary condition for the chain, the Hamiltonian in Eq.~(\ref{H0}) can be quantized as $H_{\text{SC}} = \sum_k \hbar \omega_k (a^\dag_k a_k + 1/2)$, and the charge density operator is expressed as
\begin{equation}
\vec{Q} = -i \widehat{C}^{1/2} \sum_k \vec{\psi}_k^s \sqrt{\frac{\hbar\omega_k}{2}} (a^\dag_k - a_k),
\label{Qoper}
\end{equation}
where $a_k$ ($a_k^\dag$) is the annihilation (creation) operator of mode $k$. From Eq.~(\ref{diff1}), we find that, due to the Josephson capacitances $C_J$, the equations of motion of the SQUID chain are nonlinear. In the limit $C_J \simeq 0$, the dynamical equation~(\ref{diff1}) reduces to the lumped-element model of an ordinary 1D transmission line~\cite{Gu2017s}. In this case, the capacitance matrix is simplified to
\begin{equation}
\widehat{C} \simeq \text{diag} \mleft[\ldots, C_g, C_g, C_g, \ldots \mright].
\end{equation}
Moreover, the eigenfunction $\vec{\psi}_k^s$ can be approximately written as
\begin{equation}
\vec{\psi}_k^s \simeq \sqrt{\frac{2}{N}} \mleft(\ldots, \sin\frac{kj\pi}{N}, \sin\frac{k(j+1)\pi}{N}, \ldots \mright), \quad 0 \leqslant j \leqslant N,
\end{equation}
and the charge-density operator at the antinode position in Eq.~(\ref{Qoper}) is approximately expressed as
\begin{equation}
Q \simeq -i C_g \sum_k \sqrt{\frac{\hbar\omega_k}{C_t}} (a^\dag_k - a_k), 
\label{charg_op2}
\end{equation}
where $C_t = N C_g$ is the total capacitance of the SQUID chain. In fact, to view the whole chain as a conventional 1D SQUID transmission line (STL), the condition $C_J \simeq 0$ is too strong.

In the following, we present the parameter regime where the STL has an approximately linear dispersion relation. We use the plane-wave ansatz with $\phi = A \exp{ (i\omega_{k} t - ik j d_0)}$, and by substituting it into Eq.~(\ref{diff1}), we obtain~\cite{weissl2014quantums, Krupko2018s}
\begin{equation}
\omega_k = \frac{1}{\sqrt{L_J C_g}} \sqrt{\frac{1 - \cos{(k d_0)}}{\frac{C_J}{C_g}[1 - \cos{(k d_0)}] + \frac{1}{2}}}.
\label{disreal}
\end{equation}
Moreover, we assume that the STL is approximately in the quasi-continuous regime with infinite length $L \rightarrow \infty$. Consequently, one can find that, under the conditions
\begin{equation}
d_0 \ll \lambda_k \ll L, \quad k \ll \frac{1}{d_0} \sqrt{\frac{C_g}{C_J}},
\label{LJcon}
\end{equation}
the dispersion relation is reduced to
\begin{equation}
\omega_{k0} \simeq \frac{kd_0}{\sqrt{L_J C_g}} = k v_J, \quad c_g = \frac{C_g}{d_0}, \quad l_J = \frac{L_J}{d_0},
\label{disana}
\end{equation}
where $\omega_{k0}$ is the mode frequency without $C_J$, $c_g$ ($l_J$) represents the capacitance (inductance) per unit length, and $v_J = 1/\sqrt{l_J c_g}$ is the phase velocity. Under the conditions in Eq.~(\ref{disana}), the capacitance of Josephson junctions $C_J$ can be neglected, and the wavefunction in Eq.~(\ref{diff1}) is the same as that of the discretized lumped-element circuit of a 1D transmission line.

\begin{figure*}[tbph]
	\centering \includegraphics[width=0.85\linewidth]{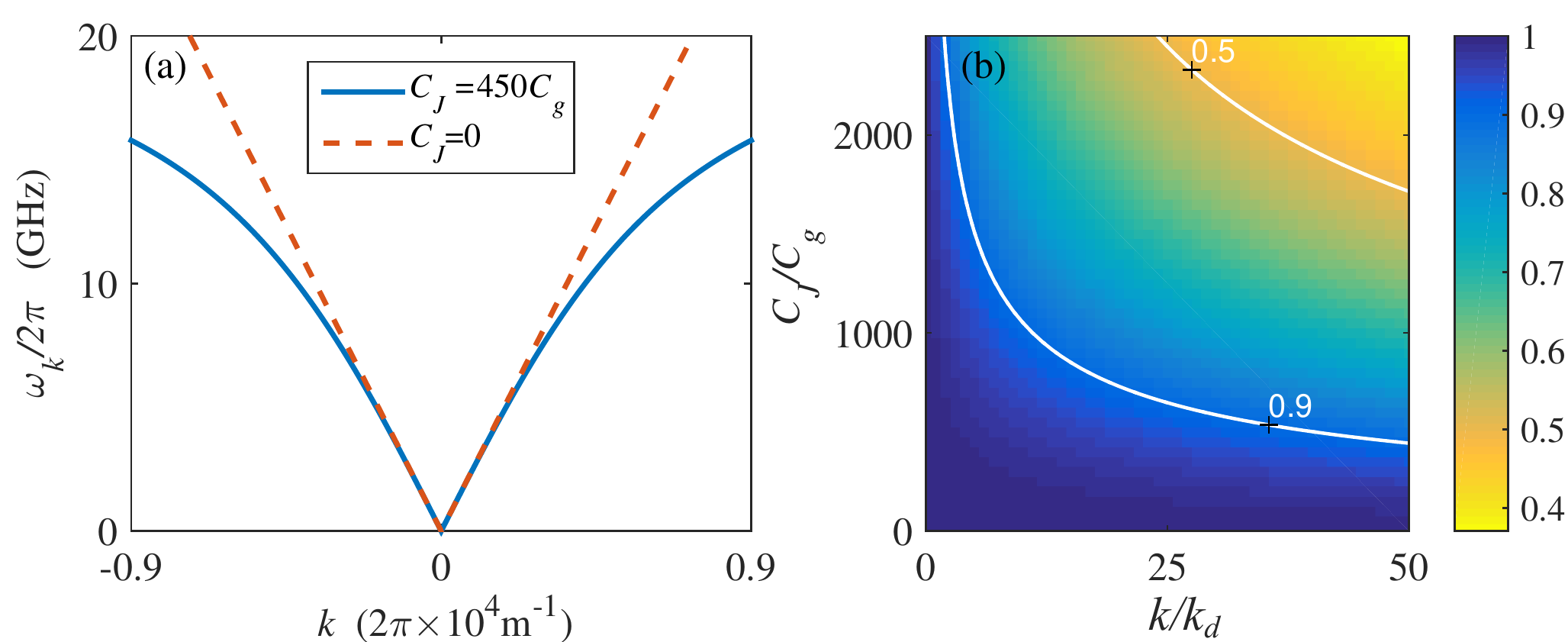}
	\caption{(a) The dispersion relation of the SQUID chain for $C_J = 0$ and $C_J = 450 C_g$, respectively. In the low-frequency limit $\omega_k < \unit[10]{GHz}$, the dispersion is approximately linear. (b) The frequency ratio $\omega_k / \omega_{k0}$ between $C_J = 450 C_g$ and $C_J = 0$ changes with mode index $k$ and Josephson capacitance $C_J$. The area delimited by the contour curve $\omega_{k}/\omega_{k0} = 0.9$ is the parameter regime where the linear dispersion relation is approximately valid. The considered SQUID number is $N = 3000$.}
	\label{fig2s}
\end{figure*}

In Table~I, we list the parameters employed in our numerical simulations. These parameters are adopted from the experimental work in Refs.~\cite{weissl2014quantums, Krupko2018s, Martinez2019s}. 
In Fig.~\ref{fig2s}(a), we plot the dispersion relation according to Eqs.~(\ref{disreal}) and (\ref{disana}), respectively. We find that, in the low-frequency regime $\omega_k/(2\pi) < \unit[10]{GHz}$, the dispersion is approximately linear even with $C_J =  450 C_g$.
In Fig.~\ref{fig2s}(b), setting $N=3000$, we numerically solve the eigenproblem in Eq.~(\ref{eigenp}) and plot the eigenfrequency ratio $\omega_k / \omega_{k0}$ as a function of Josephson capacitance $C_J$ and mode index $k$. Note that the fundamental wavevector is $k_d=2\pi/(Nd_0)$.
For nonzero $C_J$, the mode frequencies will be lower than those with $C_J = 0$. The parameter regime within the white curve $\omega_k/\omega_{k0} = 0.9$ is where $C_{J}$ will not have significant effects. The bandwidth of the deep blue area, where the STL has linear dispersion, becomes narrower when increasing $C_J/C_g$ and mode index (i.e., higher mode frequency). In the following discussions, we only focus on the parameters regime where the linear dispersion relation is valid.

\begin{table*}[tbp]
	{\normalsize \renewcommand\arraystretch{1.5}
		\begin{tabular}{>{\hfil}p{0.6in}<{\hfil}>{\hfil}p{0.5in}
				<{\hfil}>{\hfil}p{0.6in}<{\hfil}>{\hfil}p{0.5in}<{\hfil}>{\hfil}p{0.6in}<{\hfil}>{\hfil}p{1in}<{\hfil}>{\hfil}p{1.2in}<{\hfil}>{\hfil}p{1in}<{\hfil}>{\hfil}p{0.7in}<{\hfil}}
			\hline\hline  $d_0$ & $C_g$ & $C_J$ & $L_0$ & $\alpha_0$ & $\delta\alpha$ & $k_m$ & $v_J$\\
			\hline $\unit[1]{\mu m}$ & \unit[0.4]{fF} & \unit[90]{fF} & \unit[0.2]{nH} & 0.3 & 0.045 & $\unit[2\pi \times 0.3 \times 10^4]{m^{-1}}$ & $\sim \unit[10^6]{m/s}$\\
			\hline\hline
	\end{tabular}}
	\caption{The lumped-circuit parameters of the microwave PCW based on a SQUID chain that we employed for numerical simulations.} \label{table1}
\end{table*}

Compared to the standard 1D transmission line, the STL has the following advantages:
First, the characteristic impedance of the STL, $Z_R = \sqrt{l_J/c_g}$, can be much higher, which allows to realize strong coupling between superconducting atoms and STL modes~\cite{Martinez2019s}. 
Second, the impedance of each SQUID in the chain is tunable via the external flux. We can thus control the impedance of the STL via local coils, and the desired dispersion relation and exotic microwave propagating effects can be conveniently tailored for quantum optics and quantum information processing. Next, in the linear dispersion regime, we propose how to realize a PCW by periodically modulating the STL's impedance.


\section{Photonic crystal waveguides via spatially modulating the impedance}

\begin{figure}[tbph]
	\centering \includegraphics[width=0.45\linewidth]{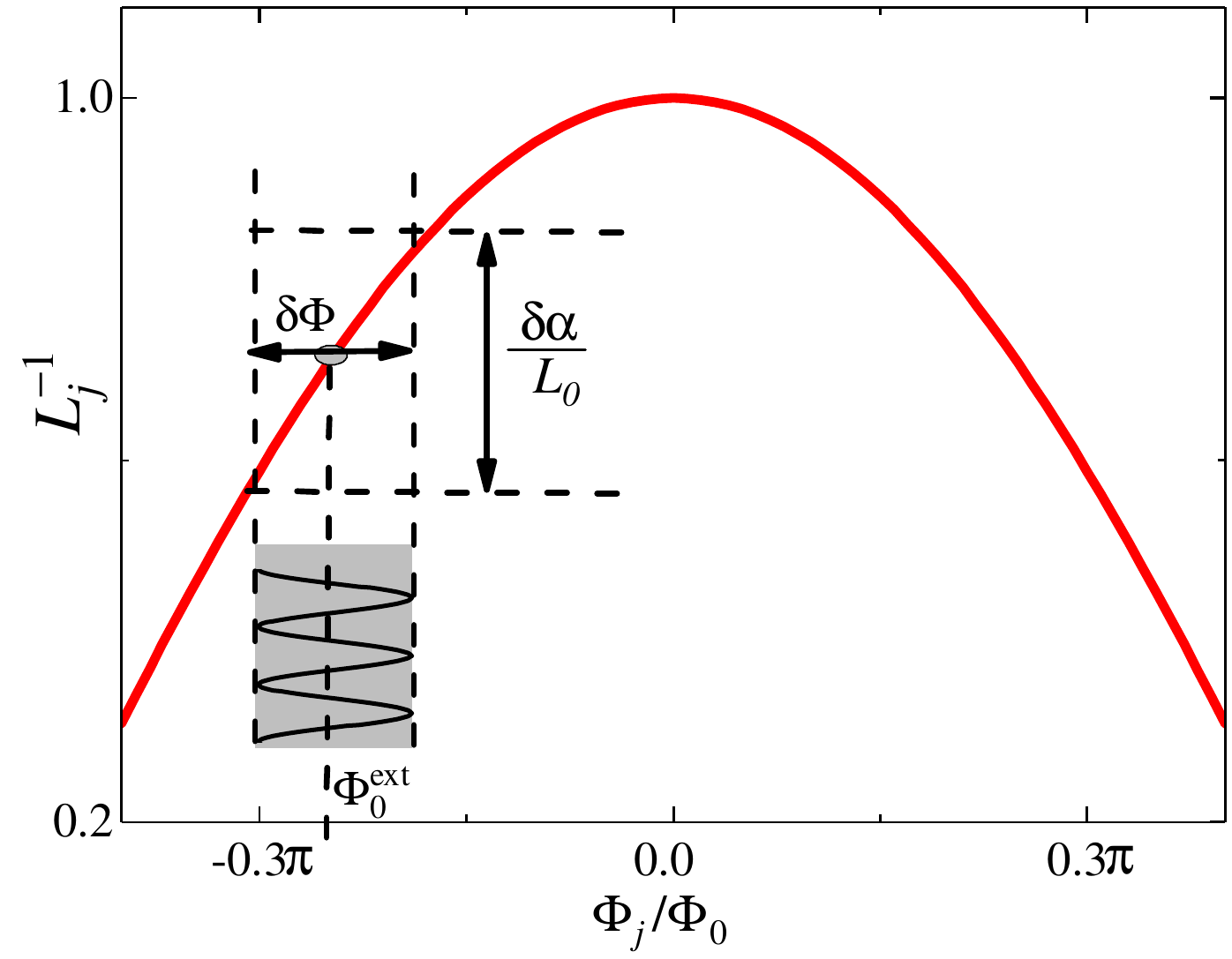}
	\caption{The parameter $L_{j}^{-1}$ changes with external flux bias around $\Phi_0^{\text{ext}}$. The photonic crystal waveguide (PCW) is realized by periodically modulating the impedance with a bandwidth $\delta\alpha/L_{0}$.}
	\label{fig3s}
\end{figure}

As depicted in Fig.~(\ref{fig1s}), to periodically modulate the STL's impedance, we consider that the flux in each SQUID loop is independently controlled by a dc modulator according to the relation
\begin{equation}
\Phi_j = \Phi_0^{\text{ext}} + \delta \Phi f (j), 
\end{equation}
where $\Phi_{0}^{\text{ext}}$ is the static flux, $\delta \Phi$ is the modulation amplitude, and $f_j$ is the position-dependent modulation signals. The modulation is depicted in Fig.~(\ref{fig3s}). We assume the STL is working as a microwave PCW, where the modulation is periodic in space. Then, the Josephson inductance can be written as
\begin{gather}
\frac{1}{L_j} \simeq \frac{1}{L_0} \left[\alpha_0 + \delta\alpha f(j) \right], \quad \alpha_0 = \cos \mleft( \frac{\pi \Phi_0^{\text{ext}}}{\Phi_0} \mright),
\label{Lmod}\\
\delta\alpha = -\sin \mleft( \frac{\pi \Phi_0^{\text{ext}}}{\Phi_0} \mright) \frac{\pi \delta \Phi}{\Phi_0}.
\end{gather} 
The inductance term in  Eq.~(\ref{diff1}) is rewritten as
\begin{eqnarray}
\frac{\phi _j - \phi _{j-1}}{L_j} - \frac{\phi_{j+1} - \phi_j}{L_{j+1}} = \frac{\phi_j - \phi_{j-1}}{L_j} - \frac{\phi_{j+1} - \phi_j}{L_j} + \frac{\phi _{j+1} - \phi_j}{L_j} - \frac{\phi_{j+1} - \phi_j}{L_{j+1}}.
\label{Linduct}
\end{eqnarray}
We assume that the distance $d_0$ between neighboring SQUIDs is much smaller than the wavelength of the field.
Therefore, by replacing $j d_0 \rightarrow x$, we use quasi-continuous functions to describe the modulation signal and fields. Consequently, we have
\begin{equation}
\phi_j (t) \rightarrow \phi(x,t), \quad f(j) = f(x).
\end{equation}
Moreover, by defining the inductance and capacitance per unit length for the STL
\begin{equation}
l(x) = \frac{L_0}{d_0}\frac{1}{\alpha_0 + \delta\alpha f(x)}, \quad c_{g,J} = \frac{C_{g,J}}{d_0},
\end{equation}
Eq.~(\ref{Linduct}) is rewritten as
\begin{equation}
\frac{\phi_j - \phi_{j-1}}{L_j} - \frac{\phi_{j+1} - \phi_j}{L_{j+1}} =
- \frac{\partial}{\partial x} \mleft[ \frac{1}{l(x)} \frac{\partial \phi(x,t)}{\partial x} \mright] d_0^{2}.
\end{equation}
Similarly, the capacitance terms in Eq.~(\ref{diff1}) can also be rewritten as a quasi-continuous function
\begin{equation}
C_g \ddot{\phi}_j + C_J \mleft( \ddot{\phi}_j - \ddot{\phi}_{j-1} \mright) - C_J \mleft( \ddot{\phi}_{j+1} - \ddot{\phi}_j \mright) = C_g \frac{\partial^2 \phi(x,t)}{\partial t^2} - C_J \frac{\partial^2 \phi(x,t)}{\partial t^2 \partial x^2} d_0^2.
\label{diffss}
\end{equation}
Therefore, in the quasi-continuous regime, Eq.~(\ref{diff1}) is written as
\begin{equation}
c_g \frac{\partial^2 \phi(x,t)}{\partial t^2} = c_J d_0^2 \frac{\partial^4 \phi(x,t)}{\partial t^2 \partial x^2} + \frac{\partial}{\partial x}\mleft[ \frac{1}{l(x)} \frac{\partial \phi(x,t)}{\partial x} \mright],
\label{waveeq}
\end{equation}
where the Josephson capacitance $c_J$ induces a nonlinear term involving both spatial and temporal differentials. 	
For simplicity, in our numerical simulations, we first consider the modulation to be on cosine form, i.e.,
\begin{equation}
\frac{1}{l(x)} = \frac{d_0}{L_0} \mleft[ \alpha_0 + \delta\alpha \cos(k_m x) \mright],
\end{equation}
where $k_m$ is the modulation wavevector. 
Consequently, the field operator $\phi(x,t)$ is written in terms of a Bloch expansion:
\begin{equation}
\phi(x,t) = e^{i(\omega_l t + k x)} u_k (x), \quad u_k (x) = \sum_{n = -\infty}^{n = \infty} c_{nk} e^{i n k_m x},
\label{phifs0}
\end{equation}
where $\omega_l$ is the eigenfrequency with $l$ the index of the energy bands, $u_k (x)$ is a spatially periodic function satisfying $u_k (x) = u_k (x + \lambda_m)$, with $\lambda_m = 2\pi / k_m$ being the period, and $c_{nk}$ is the coefficient of the $n$th Fourier order for $u_k (x)$. By substituting the wave function in Eq.~(\ref{phifs0}) into Eq.~(\ref{waveeq}), we obtain the dispersion relation between $\omega_{l}(k)$ and $k$ by solving the following quadratic eigenvalue problem:
\begin{equation}
\mleft[ \omega_l^2 (k) \hat{M}_2 + \hat{M}_0 \mright] \hat{U}(k) = 0,
\end{equation}
where 
\begin{eqnarray}
\hat{M}_2 &=& \text{diag} \mleft[\ldots, -c_J (d_0)^2 \mleft( k+n k_m \mright)^2 - c_g, \ldots \mright], \\
\widehat{M}_0 &=&
\begin{pmatrix}
\ddots & \ddots & \ddots & 0 & 0 & 0  \\
... & T_{n-1,n-2} & T_{n-1,n-1} & T_{n-1,n} &0 & 0  \\ 
... & 0 & T_{n,n-1} & T_{n,n} & T_{n,n+1} & 0  \\ 
\vdots & 0 & 0 & 0 & \ddots & \ddots
\end{pmatrix},
\end{eqnarray}
with
\begin{equation}
T_{n,n} = \frac{1}{l_0} \mleft( k + n k_m \mright)^2, \quad l_0=\frac{L_0}{\alpha_0 d_0},
\end{equation}
and
\begin{equation}
T_{n,n \pm 1} = \frac{\delta\alpha}{2 l_0} \mleft\{ \mleft( k + \mleft( n \pm 1 \mright) k_m \mright)^2 + \mleft( k + \mleft( n \pm 1 \mright) k_m \mright) k_m \mright\}.
\end{equation}
From the formulas for $\hat{M}_{2}$ and $T_{n,n}$, we find that, under the condition
\begin{equation}
c_J (d_0)^2 \mleft( k + n k_m \mright)^2 \ll c_g  \quad \longrightarrow \quad k + n k_m \ll \frac{1}{d_0} \sqrt{\frac{c_g}{c_J}},
\label{cJJ}
\end{equation}
the nonlinear terms due to the Josephson capacitance $C_J$ will not have significant effects. The condition in Eq.~(\ref{cJJ}) is similar to the condition for the linear dispersion in Eq.~(\ref{LJcon}). For the higher Fourier orders (large $n$) beyond Eq.~(\ref{cJJ}), we require that their contributions are much smaller than the lower orders. Numerical calculations indicate that by adopting small modulation amplitudes $\delta \alpha$, the coefficients $c_{nk}$ decrease quickly with Fourier order $n$.
Therefore, the nonlinear effects due to $C_J$ can be neglected. In our main text, we only consider the lowest band with $l = 1$. According to Eqs.~(\ref{charg_op2}) and (\ref{phifs0}), the charge-density operator $Q$ can be expressed with the mode operators in the first Brillouin zone (BZ)
\begin{equation}
Q(x) \simeq -i C_g \sum_{k \in \text{BZ}} \sqrt{\frac{\hbar \omega_k}{C_t}} \mleft[ a^\dag_k e^{i k x} u_k (x) - a_k e^{-i k x} u^*_k (x)\mright].
\label{chargeQ}
\end{equation}
The above charge-density operator will be employed for the coupling between the PCW and a superconducting atom.

\begin{figure}[tbph]
	\centering \includegraphics[width=0.7\linewidth]{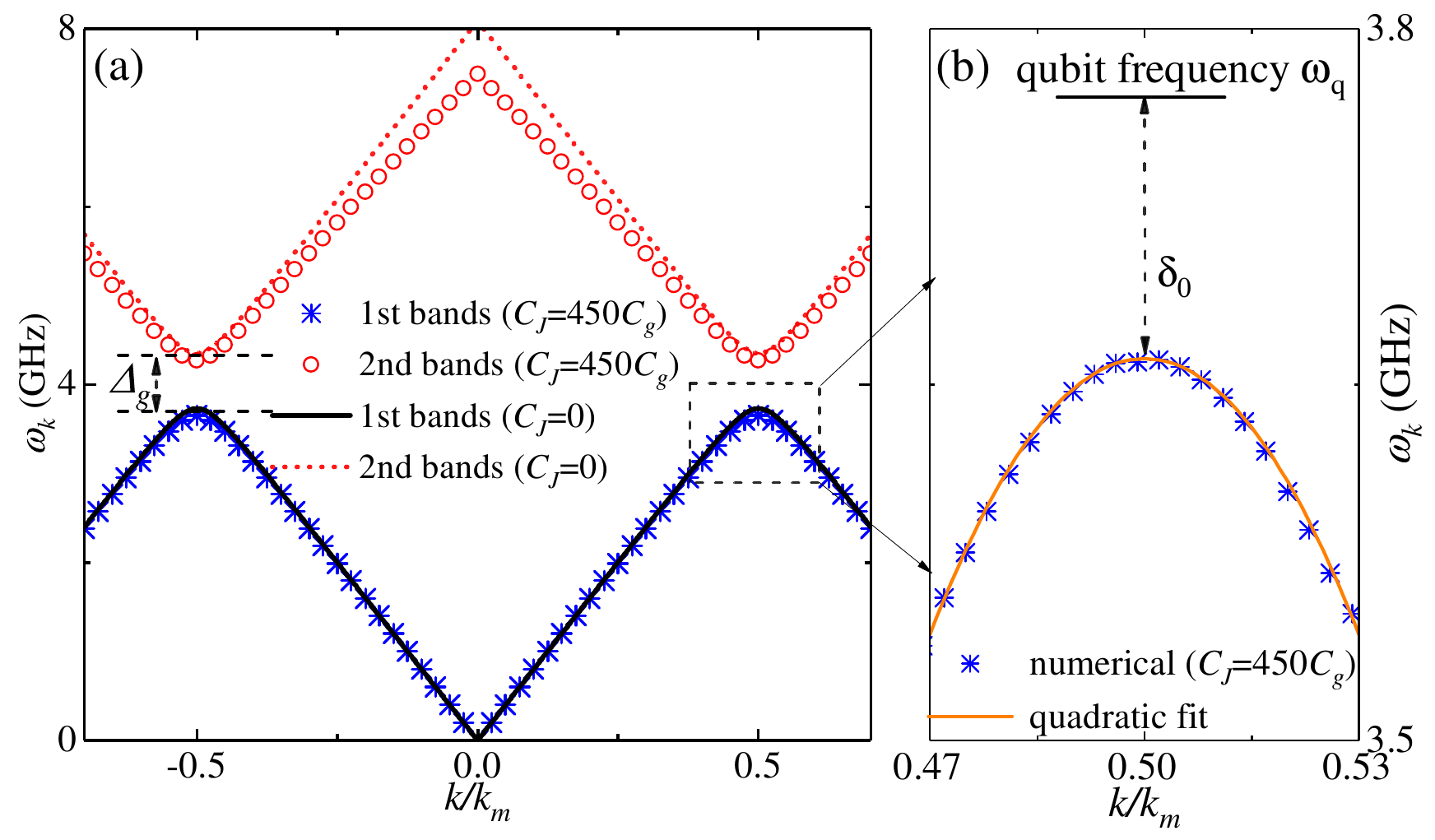}
	\caption{(a) The two lowest bands for the PCW via spatial modulation of the SQUID inductance for $C_J = 450 C_g$ and  $C_J = 0$, respectively. Parameters are taken from Table~\ref{table1}. Around $k = \pm 0.5 k_m$, there are two symmetric bandgaps with width $\Delta_g$. (b) Zoom-in around the bandgap regime. The solid curve is a quadratic fit for the dispersion relation. In our discussions, the considered atom frequency lies inside the gap with a detuning $\delta_0 = 0.1 \Delta_g$.}
	\label{fig4s}
\end{figure}

In Fig.~\ref{fig4s}(a), employing the parameters listed in Table.~\ref{table1}, we plot the band structure for the Josephson-chain PCW. We find that, even under the condition $C_J = 450 C_g$, the dispersion relations for the 1st and 2nd bands are well described by the linear approximation with $C_J = 0$. 
In the low-frequency limit, we can view the chain as a linear-dispersion medium by neglecting the Josephson capacitance under the condition in Eq.~(\ref{cJJ}). In the first Brillouin zone $k\in(-0.5k_{m}, 0.5k_{m}]$, there are two symmetric bandgaps with width $\Delta_g$ around $k = \pm 0.5 k_m$, which has been predicted in studies of 1D superconducting PCWs~\cite{Liu2017s, Sundaresan19s}. The bandgap regime is around $\omega_k / (2\pi) \simeq \unit[4]{GHz}$, which matches with the transition frequency of superconducting atoms. In Fig.~\ref{fig5s}(a), we plot the amplitudes of the Bloch wavefunctions $|u_k(x)|$ versus $x$ for the modes around the band edge. Figure.~\ref{fig5s}(b) shows the position-dependent impedance $Z(x) = \sqrt{l(x)/c_g}$ (in units of constant impedance $Z_0 = \sqrt{l_0/c_g}$) of the PCW. We find that, for the modes in the first band, $|u_k(x)|$ are highest (lowest) at the impedance dip (peak) positions, and their spatial periods all equal $\lambda_m$.

\begin{figure}[tbph]
	\centering \includegraphics[width=0.59\linewidth]{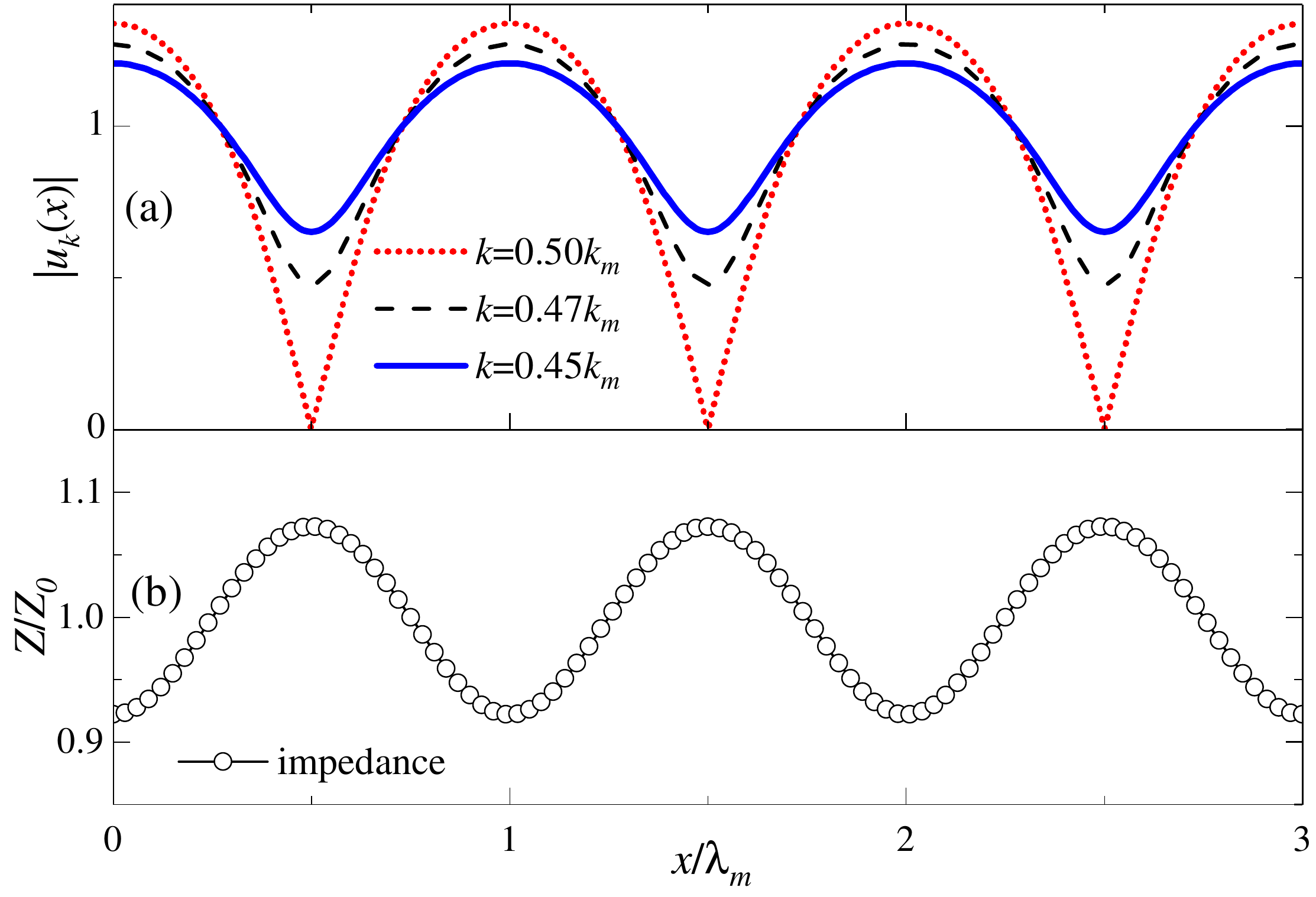}
	\caption{(a) Amplitudes of the Bloch wavefunctions $|u_k(x)|$ as a function of $x$ for the modes of the lowest band around the band edge. The impedance-modulating signal is depicted in (b). The PCW parameters are adopted from Table~~\ref{table1}.}
	\label{fig5s}
\end{figure}

In the following, we will discuss the waveguide QED for superconducting atoms interacting with the Josephson PCW.

\section{Chiral bound states induced by giant-atom effects}

The conventional interaction between cold atoms and a PCW requires optical trapping of each atom at a single position with the lowest (or highest) refractive index~\cite{Hung2013s, Chang2018s}. The natural atomic size is much smaller than the length of the PCW unit cell. In solid-state SQC systems, these limitations do not exist.
As shown in Fig.~1 of the main text, we consider a superconducting giant atom interacting with the PCW at two positions $x_{\pm}$ via capacitances $C_J^{g\pm}$. The following discussion takes the charge qubit as an example~\cite{Gu2017s}, but can also be applied for the transmon qubit~\cite{Koch07s,You2007bs}. The Hamiltonian for the superconducting atom is expressed as
\begin{equation}
H_{\text{q}} = 4E_C (\hat{n} - n_g)^2 - 2 E^q_J \cos \mleft( \frac{\pi \Phi_{\text{q}}}{\Phi_0} \mright) \cos{\phi},
\end{equation}
where $E_C = e^2/(2C_\Sigma)$ is the charging energy of the atom's junctions, $C_\Sigma = C^q_J + C_J^{g-} + C_J^{g+}$, with $C^q_J$ the Josephson capacitance, and $E^q_J$ is the Josephson energy of one junction in the atom. Note that $\Phi_{\text{q}}$ is the control flux through the split junction's loop. This flux is employed for tuning the atom's transition frequency. Around the charge degeneracy point $n_g = 1/2$, the above Hamiltonian can be quantized in a qubit basis as
\begin{equation}
H_{\text{q}} = - E^q_J \cos \mleft( \frac{\pi\Phi_{\text{q}}}{\Phi_0} \mright) \mleft( |0\rangle\langle1| + |1\rangle\langle0| \mright) - 4 E_C \delta n_g \mleft( |1\rangle\langle1| - |0\rangle\langle0| \mright).
\end{equation}
The offset-charge deviation $\delta n_{g}$ is written as
\begin{equation}
\delta n_{g} = \sum_{\pm} 
\frac{Q(x_{\pm})}{C_g} \frac{C_J^{g\pm}}{2e} = -i \sum_{\pm} \sum_{k \in \text{BZ}} \frac{C_J^{g\pm}}{2e} \sqrt{\frac{\hbar \omega_k}{C_t}} \mleft[ a^\dag_k e^{i k x_{\pm}} u_k (x_{\pm}) - a_k e^{-i k x_{\pm}} u^*_k (x_{\pm})\mright],
\label{chargen}
\end{equation}
where $Q(x_{\pm})$ is the charge-density operator at two coupling positions $x_{\pm}$ described by Eq,~(\ref{chargeQ}).
In the basis
\begin{equation}
|e\rangle=\frac{|1\rangle-|0\rangle}{\sqrt{2}}, \quad  |g\rangle=\frac{|1\rangle+|0\rangle}{\sqrt{2}},
\end{equation}
the Hamiltonian for this coupled circuit-QED system is written as
\begin{equation}
H_0 = \frac{1}{2} \hbar \omega_q \sigma_z + \sum_k \hbar \omega_k a^\dag_k a_k + i \sum_k \hbar \mleft( g_k a^\dag_k - g^*_k a_k \mright) (\sigma_+ + \sigma_-),
\label{HO}
\end{equation}
where
\begin{equation}
\omega_q = \frac{2 E^q_J}{\hbar} \cos \mleft( \frac{\pi \Phi_{\text{q}}}{\Phi_0} \mright)
\end{equation}
is the atomic transition frequency. The giant-atom coupling strength with mode $k$ is
\begin{gather}
g_k = \sum_{i=\pm} g_k^i e^{i k x_i} u_k (x_i), \quad 
g_k^\pm = \frac{e}{\hbar} \frac{C_J^{g \pm}}{C_\Sigma} \sqrt{\frac{\hbar \omega(k)}{C_t}} \simeq \frac{e}{\hbar} \frac{C_J^{g \pm}}{C_\Sigma} \sqrt{\frac{\hbar \omega_q}{C_t}},
\label{gkG}
\end{gather}
where the mode frequency $\omega(k)$ is approximately replaced by the qubit frequency $\omega_q$. Consequently, $g_k^\pm$ will approximately become independent of $k$. 
Note that Eq.~(\ref{gkG}) is derived by assuming the impedance of the STL, $Z_J = \sqrt{L_J / C_g}$ is much smaller than the impedances of the coupling capacitance and the superconducting atom, i.e.,
\begin{equation}
Z_J \ll \max\{(\omega_q C_J^g)^{-1}, Z_{q} \},
\label{ZJcon}
\end{equation}
where $Z_q$ is the characteristic impedance of the atom, which can be estimated from its lumped-circuit model~\cite{Masluk2012s}. In this case, we can view the STL as a low-impedance environment. However, compared with the conventional transmission line with character impedance $Z_0 \simeq \unit[50]{\Omega}$, $Z_J$ can be much larger, and enables the realization of strong coupling between a superconducting atom and STL modes~\cite{Altimiras2013s, Martinez2019s}. 
For example, employing the parameters in Table~I, the estimated STL impedance is about $Z_J \simeq \unit[550]{\Omega}$. 
When the characteristic impedances of the superconducting atom and the STL match up with $Z_J \sim Z_q$, the system enters into the overdamped regime, with the coupling strength reaching its maximum value~\cite{Martinez2019s}. Consequently, the coupling form in Eq.~(\ref{gkG}) will be significantly modified.
Therefore, to satisfy the impedance relation in Eq.~(\ref{ZJcon}), the coupling capacitance should be smaller than that employed in the standard 1D transmission line, together with the atom working as a high-impedance circuit element.

As shown in Fig.~\ref{fig4s}, we assume that the qubit transition frequency $\omega_q$ is close to the first band, and the detuning  $\delta_0$ from the band edge is much smaller than the bandgap width $\Delta_g$. Therefore we can approximately consider only the contributions of the first band. In our discussions here, we set $\delta_0 \simeq 0.1 \Delta_g$.
In a frame rotating with $\omega_q$, adopting the rotating-wave approximation, the Hamiltonian in Eq.~(\ref{HO}) becomes (setting $\hbar=1$)
\begin{equation}
H_{\text{int}} = \sum_{k \in \text{BZ}} \Delta_k (a^\dag_k a_k) 
+ \sum_{k \in \text{BZ}} (g_k a^\dag_k \sigma_- + g^*_k a_k \sigma_+),
\label{Hrtot}
\end{equation}
where $\Delta_k = \omega_k - \omega_q$ is the frequency detuning. We first define the spatial field operator expanded in terms of the Bloch wavefunctions
\begin{equation}
\phi^{\dagger}(x) = \frac{1}{\sqrt{L}} \sum_{k \in \text{BZ}} a_k^\dag e^{ikx} u_k (x),
\label{phifs}
\end{equation}
where $\phi^\dag (x)$ [$\phi (x)$] represents creating (annihilating) a photon at position $x$ and satisfies $[\phi(x), \phi^\dag (x')] = \delta(x-x')$. The bound state of the system is the eigenstate for $H_{\text{int}}$ with eigenenergy $\epsilon_{b}$, i.e., $H_{\text{int}}|\psi_{b}\rangle = \epsilon_b |\psi_b \rangle$. In the single-excitation subspace, $|\psi_b \rangle$ is
\begin{equation}
|\psi_b \rangle = \cos(\theta)|e,0\rangle + \sin \theta \sum_k c_k  \ket{g,1_k},
\end{equation}
The solution for the bound state reads
\begin{eqnarray}
c_k &=& \frac{g_k}{\tan\theta(\epsilon_b - \Delta_k)}, \label{ckk} \\
\epsilon_b &=& \sum_{k\in\text{BZ}} \frac{|g_k|^2}{(\epsilon_b - \Delta_k)}, \\
\tan \theta &=& \sum_{k\in\text{BZ}} \frac{|g_k|^2}{(\epsilon_b - \Delta_k)^2}.
\end{eqnarray}
We consider the conventional case where most of the energy of the excitation is localized in the atom, while the photonic modes are weakly populated~\cite{Douglas2015s, Liu2017s}.
In this case, $\cos(\theta) \simeq 1$ and $\epsilon_b \simeq 0$. Consequently, the wavefunction $\phi_b (x)$ of the photonic part in the PCW is
\begin{eqnarray}
\phi_b (x) &=& \sin \theta\langle x|\sum_{k\in\text{BZ}} c_k a_k^\dag|0\rangle \notag \\
&=& \sum_{k\in\text{BZ}} \frac{c_k \sin \theta}{\sqrt{L}}
\int dx' \langle x|e^{-ikx'} u_k^* (x') \phi^\dag (x')|0\rangle.
\label{phibound01}
\end{eqnarray}
By substituting $c_k$ [Eq.~(\ref{ckk})] into Eq.~(\ref{phibound01}), we obtain
\begin{equation}
\phi_b (x) \simeq \frac{\sqrt{L}}{2\pi} \int_{k\in\text{BZ}} \frac{g_k u_k^* (x) e^{-ikx}}{\epsilon_b - \Delta_k} dk,
\label{phibound0}
\end{equation}
where the integration is limited to the first BZ. As shown in Fig.~2(c) in the main text, around the band edge $k_0 \simeq k_m/2$, the real part of $g_{k}$ is approximately constant. However, the imaginary part is not constant, but changes with $\delta k = k - k_0$ rapidly and linearly, which is completely different from the small-atom case. Therefore, we should write
\begin{equation}
g_k \simeq (A + i B \delta k),
\label{gkri}
\end{equation}
where $A$ is the average of the real part for $g_k$ around $k_0$ and $B$ is the slope of the imaginary part of $g_k$ changing with $k$. For giant atoms, B is non-zero. Around the band edge of the PCW, we use the effective-mass approximation by expanding the dispersion relation as a parabolic function~\cite{Douglas2015s, GonzlezTudela2015s}. As depicted in Fig.~\ref{fig4s}(b), the dispersion relation of the PCW is well described by a quadratic function, i.e., $\Delta_k = -\delta_0 - \alpha_m (k - k_0)^2$.

Finally, we obtain
\begin{gather}
\phi_b (x) \simeq A_m \sum_{\pm} \int_{-\infty}^{\infty} d\delta k \mleft[ \frac{ C_{\pm} e^{-i\delta kx}}{\sqrt{2\pi}(\sqrt{\frac{\delta_0}{\alpha_m}} \mp i \delta k)} \mright] \label{phibound}, \\
A_m = \frac{\sqrt{L} u_{k0}^* (x) e^{-ik_{0} x}}{2\sqrt{2\pi \alpha_m \delta_0}},
\end{gather}
where $A_m$ is the amplitude for the bound state's photonic part, and $C_{\pm}$ are determined by the behavior of the imaginary and real parts of $g_k$:
\begin{equation}
C_{\pm} = A \pm B \sqrt{\frac{\delta_0}{\alpha_m}}.
\label{cpm}
\end{equation}
By integrating Eq.~(\ref{phibound}), we obtain
\begin{equation}
\phi_b (x) = A_m \mleft[ C_- \Theta(-x) + C_+ \Theta(x) \mright] \exp \mleft(-\frac{|x|}{L_{\text{eff}}} \mright),
\label{phib_ana}
\end{equation}
where $L_{\text{eff}} = \sqrt{\alpha_m/\delta_0}$ is the length scale determining the exponential decay of the localized bound state with distance, which is similar to previous studies~\cite{Douglas2015s,Hood2016s,Douglas2016s}.
Moreover, during the derivation of Eq.~(\ref{phib_ana}) we assume $|x_+ - x_-| < \lambda_m \ll L_{\text{eff}}$. When considering the bound-state distribution, we have $x_+ \simeq x_- = 0$. Therefore, the photonic energy localized between two coupling points can be neglected.

\section{The interference mechanism of the bound states in giant atoms}
\begin{figure}[tbph]
	\centering \includegraphics[width=0.9\linewidth]{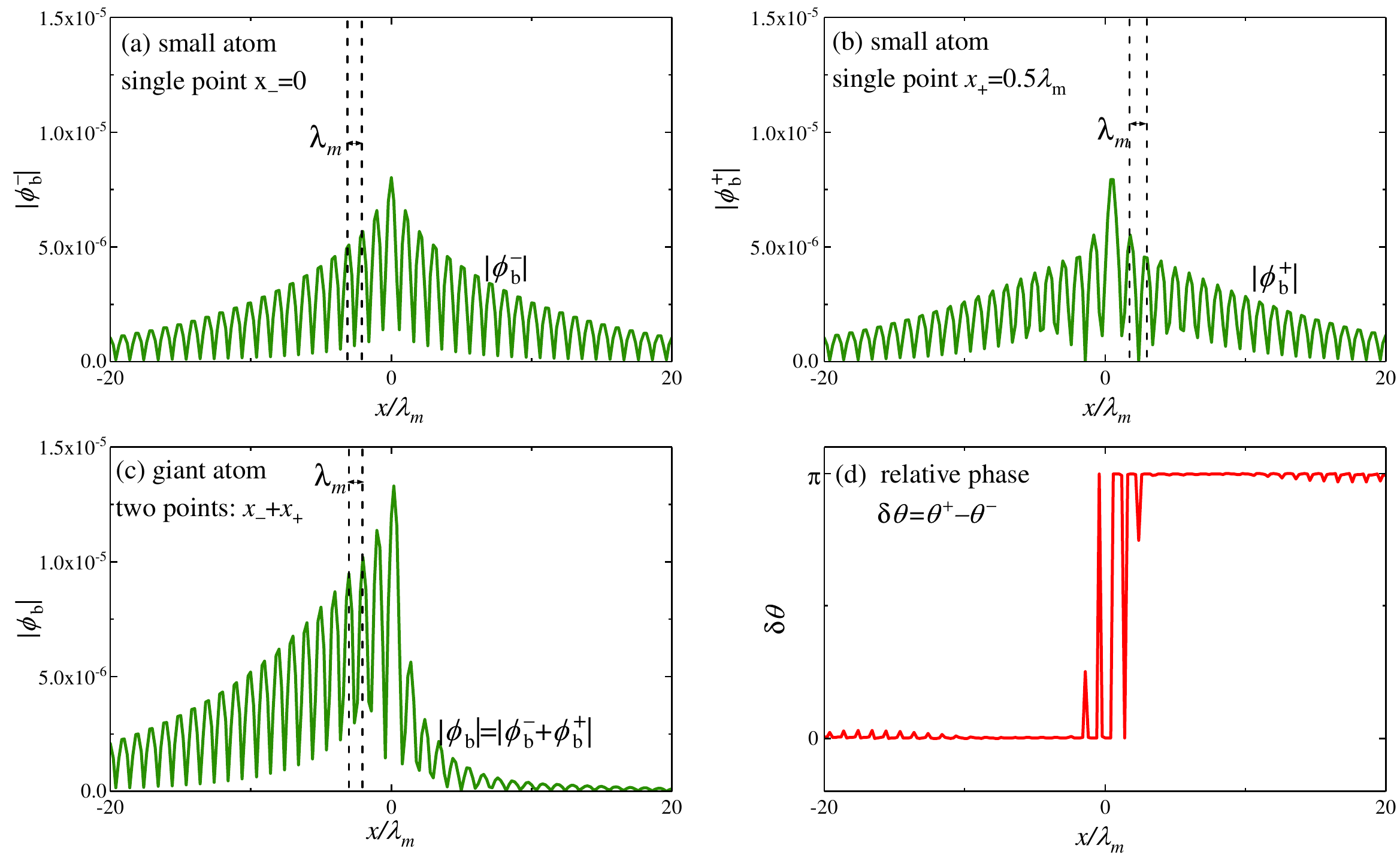}
	\caption{By setting $\{x_-, x_+\} = \{0, 0.5 \lambda_m \}$ and $g_k^+ \simeq 3.4 g_k^-$, the bound-state components (a) $\phi^-_b$, (b) $\phi^{+}_{b}$, (c) the bound state $\phi_{b}$, and (d) the phase difference $\delta \theta (x)$, change with position $x$. The PCW parameters are adopted from Fig.~\ref{fig4s}.}
	\label{fig6s}
\end{figure}

When considering a giant atom, the bound-state distribution in Eq.~(\ref{phib_ana}) is significantly affected by the interference effects between different coupling points.
To verify this, we can rewrite the bound state in Eq.~(\ref{phibound0}) as
\begin{gather}
\phi_b (x) = \phi_b^+ (x) + \phi_b^- (x), \notag \\
\phi^{\pm}_b (x) \simeq \frac{\sqrt{L}}{2\pi} \int_{k \in \text{BZ}} \frac{g_k^{\pm} e^{i k x_{\pm}} u_k (x_{\pm}) u_k^* (x) e^{-i k x}}{\epsilon_b - \Delta_k} dk = A^{\pm}_b (x) e^{i \theta_{\pm}(x)},
\label{BSinter}
\end{gather}
where $\phi^{\pm}_b (x)$ are the bound states induced by a small atom coupling at the single position $x_{\pm}$, and
$A^{\pm}(x)$ [$\theta_{\pm}(x)$] are their amplitudes (phases), which are both position-dependent.
Equation~(\ref{BSinter}) indicates that the total bound state $\phi_b (x)$ is the result of interference effects, and is determined by the phase difference $\delta \theta(x)=\theta_{+}(x)-\theta_{-}(x)$.

In Fig.~2 of the main text, by considering $x_-$ ($x_+$) at the lowest (highest) impedance position (i.e., $\{x_-, x_+ \} = \{0,  0.5 \lambda_m \}$), we discuss the bound-state behavior affected by the interference effects. In numerical discussions,
the PCW parameters are adopted from the experimental data in Table~\ref{table1}, and the atom frequency is assumed to be inside the gap with a detuning $\delta_0 \simeq 0.1 \Delta_g$.
As depicted in Fig.~\ref{fig6s}(a, b), both $\phi^+_b (x)$ and $\phi^-_b(x)$ show no chirality. However, their phase difference is approximately described by 
$\delta\theta \simeq \pi\Theta(x)$, with $\Theta(x)$ the Heaviside step function [see Figure~\ref{fig6s}(d)], indicating that the interference is constructive (destructive) in the direction $x<0$ ($x>0$). 
By setting $g_k^+ \simeq 3.4 g_k^-$, we find that $A^+ (x) \simeq A^- (x)$.
Therefore, the bound state of the giant atom is strongly localized in the left part. On the right-hand side, the bound state is mostly cancelled by the destructive interference [see Figure~\ref{fig6s}(d)]. Note that the oscillating amplitudes of the bound states are due to the periodic Bloch wavefunctions. As shown in Eq.~(\ref{phibound0}), all the periodic modes $u_k(x)$ around the band edge will contribute to the bound states $\phi_b(x)$ and $\phi^\pm_b(x)$. According to Fig.~\ref{fig5s}, $u_k(x)$ has the same period as $\lambda_m$. Therefore, the amplitude of the bound state rapidly oscillates on the scale of the decay length $L_{\text{eff}}$, which is much larger than the length of the PCW unit cell. For both giant and small atoms, the distance between two peaks in the bound states is also equal to $\lambda_m$ (see Fig.~\ref{fig6s}).

\begin{figure}[tbph]
	\centering \includegraphics[width=0.9\linewidth]{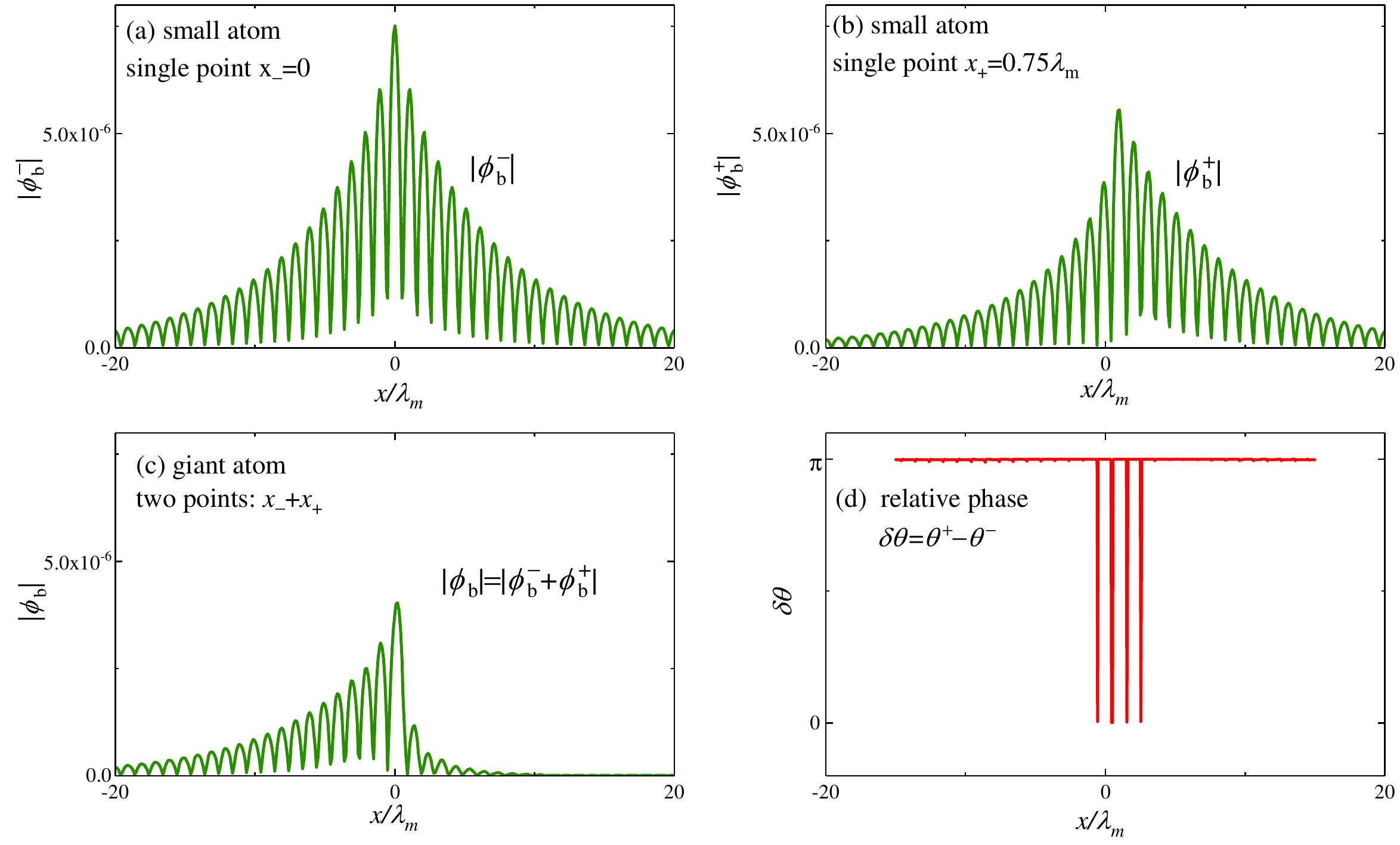}
	\caption{By setting $\{x_-, x_+ \} = \{0, 0.75 \lambda_m \}$ and $g_k^- = g_k^+$, the bound-state components (a) $\phi^-_b$, (b) $\phi^{+}_{b}$, (c) the bound state $\phi_{b}$, and (d) the phase difference $\delta \theta (x)$, change with the position $x$. The PCW parameters are adopted from Fig.~\ref{fig4s}.}
	\label{fig7s}
\end{figure}

When considering the second coupling point shifted to $x_+ = 0.75 \lambda_m$, we find another interference pattern affecting the chirality of the bound state. In Fig.~\ref{fig7s}, we plot $\phi^{\pm}_b (x)$, $\phi_b (x)$ and $\delta\theta$ as a function of $x$. The bound state $\phi^+_b (x)$ is slightly chiral due to breaking the mirror symmetry of the PCW [see Fig.~\ref{fig7s}(b)].
However, the chirality is not large. As shown in Fig.~\ref{fig7s}(d), the phase difference $\delta\theta$ is approximately equal to $\pi$ when $|x|\gg 0$. Therefore, the interference is always destructive. The amplitude for the bound state $\phi_b (x)$ of the giant atom is approximately $A_b (x) = A^-_b (x) - A^+_b (x)$.
Under the condition $g_k^+ \simeq g_k^-$, we have the relations
\begin{eqnarray}
A_b (x) = A^-_b (x) - A^+_b (x) \simeq 0, \qquad x&>&0, \notag \\
A_b (-x) \gg A_b (x)\simeq 0,  \qquad x&>&0,
\end{eqnarray}
which indicate that the bound state is strongly localized on the left side due to the quantum interference. In this case, the quantum interference effect significantly enhances the bound-state chirality.

As shown in Fig.~\ref{fig7s}, due to the destructive interference effects, the photonic energy of the bound state $\phi_b (x)$ is suppressed and smaller than $\phi^{\pm}_b (x)$. Similar to optical interference, we can define the interference visibility of the bound state as
\begin{equation}
W = \frac{\int_{-\infty}^{\infty}dx|\phi_b(x)|^2}{\int_{-\infty}^{\infty}dx|\phi^+_b (x)|^2 + \int_{-\infty}^{\infty}dx|\phi^-_b (x)|^2},
\end{equation}
from which one finds that $W = 0$ ($W = 2$) indicates that the interference is maximally destructive (constructive), and the bound state vanishes (is enhanced).

In Fig.~\ref{fig8s}(a), setting $x_- = 0$ and $g_k^- = g_k^+$, we plot the interference visibility $W$ as a function of $x_+$. We find \emph{another unconventional behaviour} of the bound state: when the separation distance satisfies $$d_g = x_+ - x_- = (2N + 1) \lambda_m$$ with $N$ integer, $W\simeq 0$, indicating that the bound state is completely cancelled.
The mechanism for the disappearance of the bound state can be understood as follows: only the modes around the band edge contribute significantly to the bound state. In the coupling formula in Eq.~(\ref{gkG}), we approximately replace $k$ with $k_m/2$. Therefore, for mode $k$, we can write
\begin{equation}
e^{ikx_+} u_k (x_+) = e^{i k (x_- + d_g)} u_k (x_- + d_g) \simeq \left\{
\begin{array}{lr}
-e^{ikx_-} u_k (x_-), \quad d_g = (2N+1) \lambda_m, &  \\
+e^{ikx_-} u_k(x_-), \quad d_g = 2N \lambda_m, & 
\end{array}
\right.
\label{nlambda}
\end{equation}
where we have employed the properties of the Bloch wavefunctions $u_k (x_- + N \lambda_m) = u_k (x_-)$, and $k_m \lambda_m / 2 = \pi$.
Equation~(\ref{nlambda}) indicates that, when $d_g = (2N+1) \lambda_m$, the interference between $\phi^{\pm}_b (x)$ will cancel the two contributions completely, leading to $\phi_b (x) \simeq 0$, i.e., the bound state vanishes completely.
Conversely, at positions $d_g = 2N\lambda_m$ the interference is maximally constructive with $W \simeq 2$.

\begin{figure}[tbph]
	\centering \includegraphics[width=0.8\linewidth]{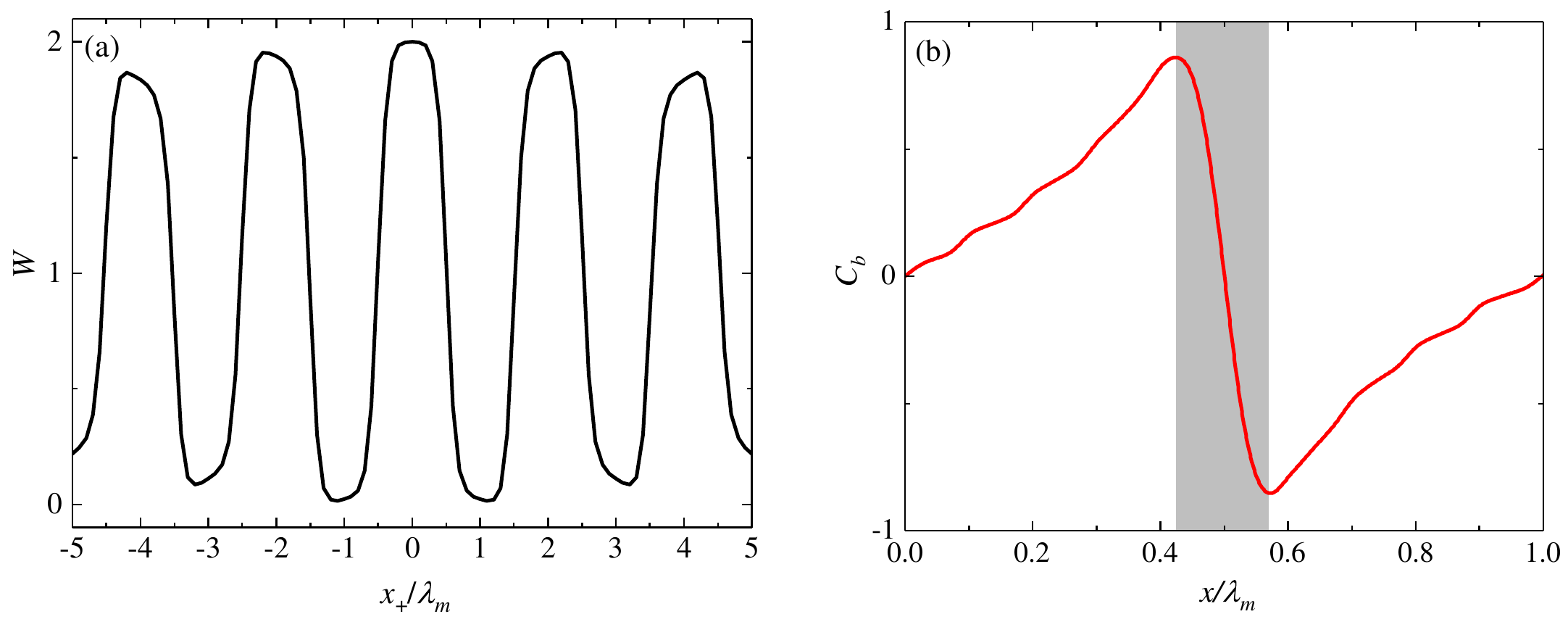}
	\caption{(a) Setting $x_- = 0$ and $g_k^- = g_k^+$, the interference visibility $W$ changes with the second coupling position $x_+$. The PCW parameters are adopted from Fig.~\ref{fig4s}. (b) The bound-state chirality changes with the coupling position of a small atom.
	}
	\label{fig8s}
\end{figure}

If a small SQC atom does not couple to the lowest (or highest) impedance position, it can see different semi-infinite waveguide structures in different directions if we split the PCW into two halves at the single coupling point. As shown in Fig.~\ref{fig7s}(b), 
the bound state of a small atom already shows chiral behaviour given that the coupling position is at $x_+ = 0.75 \lambda_m$. 
In Fig.~\ref{fig8s}(b), considering a small SQC atom, we plot the chirality $\mathcal{C}_b$ [defined in Eq.~(8) in the main text] as a function of the coupling position, and find that \emph{the chirality changes rapidly} around the highest impedance position $x = 0.5 \lambda_m$.
The chirality for the small atom is \emph{not} due to the quantum interference effects discussed for giant atoms.
In a narrow regime $x\in[0.43\lambda_m,0.57\lambda_m]$ (grey area),  the bound state varies from close to maximally left to close to maximally right chirality, indicating that the coupling position has to be fixed accurately to achieve a certain chirality. In the giant-atom case, as depicted in Fig.~3 of the main text, the opposite chiral relations occur only when $x_+$ is located in the opposite direction of $x_-$, with a much larger separation distance.
Moreover, in small-atom systems, the chirality cannot be tuned by changing the coupling strength, while in giant-atom systems, \emph{the chirality can be continuously changed} by modulating the relative giant-atom coupling strengths [see Fig~3(b) in the main text]. In conclusion, compared to the small atom, the chirality in giant atom system is due to a different mechanism, and is more flexible in experimental implementations.


\section{Chiral dipole-dipole interactions mediated by virtual photons}
\subsection{Chiral dipole-dipole interactions}
Here we derive the chiral dipole-dipole interactions between multiple atoms induced by the giant-atom effects. We assume that all atomic transition frequencies are identical, $\omega_q$. In a frame rotating with $\omega_q$, the Hamiltonian of the whole system reads
\begin{equation}
H_0^{\text{m}} = \sum_{k \in \text{BZ}} \Delta_k a_k^\dag a_k + \sum_i \sum_{k \in \text{BZ}} (g_{ki} \sigma_i^- a_k^\dag + \text{H.c.}),
\label{Hmq}
\end{equation}
where $g_{ki}$ is given in Eq.~(\ref{gkG}). As depicted in Fig.~4 of the main text, we first consider the intracell coupling ($i=A,B$). Since the modes $\pm k$ are degenerate with $\omega(k) = \omega(-k)$, we restrict $0 < k^+ < k_m/2$ in the positive BZ. The coupling strengths satisfy $g_{ki}^* = g_{-ki}$. The atomic operators can be written in the symmetric and antisymmetric forms as $S_{\pm} = (\sigma_A^- \pm \sigma_B^-) / \sqrt{2}$.
Moreover, we define the supermode operator of the bath modes as
\begin{equation}
a_{k,\pm} = \frac{(g_{kA}^* \pm g_{kB}^*) a_k + (g_{kA} \pm g_{kB}) a_{-k}}{\sqrt{2}|g_{kA}\pm g_{kB}|}, 
\end{equation}
where the commutation relation satisfies
\begin{equation}
[a_{k,\beta}, a_{k',\beta'}] = \delta_{kk'} \delta_{\beta\beta'} \delta(|g_{kA}| - |g_{kB}|).
\label{orthogonal}
\end{equation}
which indicates that, under the condition $|g_{kA}| = |g_{kB}|$, the symmetric and antisymmetric operators $S_{\pm}$ are coupled to independent baths $a_{k,\pm}$, and their evolutions are separable~\cite{GonzlezTudela2017s}. Therefore, the interaction Hamiltonian in Eq.~(\ref{Hmq}) is rewritten as
\begin{equation}
H_{0}^{\text{m}} = \sum_{k^+, \beta = \pm} \mleft[ \Delta_k a_{k, \beta}^\dag a_{k, \beta} + G_k^\beta (S_\beta a_{k, \beta}^\dag + \text{H.c.})\mright],
\label{supermH}
\end{equation}
where
\begin{equation}
G_k^{\pm} = |g_{kA} \pm g_{kB}|
\end{equation}
are the coupling strengths between $S_{\pm}$ and the supermodes $a_{k, \pm}$. Note that $\Delta_k = \omega_k - \omega_q$ in Eq.~(\ref{supermH}) is kept unchanged but only limited by $k^+ > 0$. We denote the initial states as $|\Psi_{\pm}\rangle = S_{\pm}|g,g,0\rangle$, where $|g(e)\rangle$ and $|0\rangle$ represent the qubit in the ground (excited) state and the PCW in the vacuum state, respectively. Using standard resolvent-operator techniques~\cite{cohen1998atoms}, the probability amplitudes $C_{\pm}(t)$ ($t>0$) that the whole system remains in $|\Psi_{\pm}\rangle$ are derived as
\begin{equation}
C_{\pm}(t) = \frac{i}{2\pi} \int_{-\infty}^{\infty} dE G_{\pm} (E+i0^{+})e^{-iEt},
\label{Laplace}
\end{equation}
where $G_{\pm}(z)$ are the retarded Green functions~\cite{cohen1998atoms}, and $z = E + i0^{+}$ is Fourier frequency above the real axis. Given that $|g_{kA}| = |g_{kB}|$, $G_{\pm}(z)$ are expressed in simple analytical forms as
\begin{gather}
G_{\pm}(z) = \frac{1}{z - \Sigma_e (z) \mp \Sigma_{AB}(z)}, \\
\Sigma_e (z) = \int_0^{k_0} dk \frac{2(|g_{kA}|^2 + |g_{kB}|^2)}{z - \Delta_k}, \\
\Sigma_{AB} (z) = \int_0^{k_0} dk \frac{2 \Re(g_{kA} g_{kB}^*)}{z - \Delta_k},
\label{sig12}
\end{gather}
where
$\Sigma_e (z) \mp \Sigma_{AB} (z)$ is the atomic self-energy that describes the coupling effect between the atoms and PCW modes. In our discussions, we always assume that $|g_{kA}| \simeq |g_{kB}|$.
When the coupling strengths $|g_{kA}|$ and $|g_{kB}|$ differ by a lot, the orthogonality condition of the modes $a_{k,\pm}$ in Eq.~(\ref{orthogonal}) is not valid~\cite{GonzlezTudela2017s}. Consequently, there is a tunnelling term $(a_{k,+}^\dag a_{k,-} + \text{H.c.})$ between two baths, which describes the entangled  evolutions between states 
$|\Psi_{\pm}\rangle$. In this case, the energy denominators for the Green functions $G_{\pm}(z)$ become much more complicated. 

By assuming that the giant-atom couplings are sufficiently weak~\cite{cohen1998atoms}, the standard Born-Markov approximation is valid, and we can replace $E$ as the atom frequency. We then approximately obtain
\begin{eqnarray}
\text{Re}[\Sigma_e (z)] \simeq \text{Re}[\Sigma_e (\omega_q + i0^+)] = \delta_{qs}  \\
\text{Re}[\Sigma_{AB} (z)] \simeq \text{Re}[\Sigma_{AB} (\omega_q + i0^+)] = J_{AB},
\label{firstoder}
\end{eqnarray}
where the imaginary parts of $\Sigma_e (z)$ and $\Sigma_{AB} (z)$, describe the individual and collective decay of the atoms, respectively. Since the atoms interact with the PCW bandgap, in next section we show that the decay effects are strongly suppressed with a large detuning $\delta_0$.
Note that $\delta_{qs}$ represents the vacuum Stark shift of the atoms due to coupling with the PCW modes, which is the same for states $|\Psi_{\pm}\rangle$ [see Fig.~\ref{fig10s}(a)]. The important quantity is $J_{AB}$, the real part of $\Sigma_{AB} (z)$, which is in fact equal to the Rabi frequency
of the coupling between states $|g,e,0\rangle$ and $|e,g,0\rangle$, and describes the coherent dipole-dipole coupling mediated by virtual photons in the PCW~\cite{Douglas2015s, GonzlezTudela2017s}. As discussed in the main text, even when the atoms are equally spaced, due to giant-atom-induced interference effects, the coupling strengths show chiral preference with $J_{AB} \neq J_{BA}$.

Finally, we discuss the effects of different kinds of impedance modulation signals on our proposal. We consider the following square and cosine modulation signals:
\begin{equation}
\frac{1}{l(x)}=\left\{
\begin{array}{lr}
\frac{d_0}{L_0} \mleft[ \alpha_0 + \delta\alpha \cos(k_m x) \mright], \quad ~~~~~~~~~~~~~\text{cosine wave}, & \\
\frac{d_0}{L_0} \mleft\{\alpha_0 + \delta\alpha \ \text{sgn}[\cos(k_m x) ]\mright\} \quad ~~~~~~\text{square wave}, & 
\end{array}
\right.
\end{equation} 
where sgn is the signum function. We plot the chirality $\mathcal{C}_b$ versus $x_{+}$ in Fig.~\ref{fig9s}(a). In spite of the quite different shapes of the modulation signals, their chiralities remain nearly the same. In Fig.~\ref{fig9s}(b), we show the dipole-dipole interaction strengths $J_{AB}$ and $J_{BA}$ as a function of the separation $D_q$ of two atoms.  Relative to the case for the cosine modulation, the coupling strengths decay a little faster for the square-wave modulation due to its smaller decay length $L_{\text{eff}}$. However, by comparing $J_{AB}$ and $J_{BA}$, we can infer that the dipole-dipole interactions remain chiral for both impedance modulations. Therefore, our proposal is insensitive to the shape of modulation signals.

\begin{figure}[tbph]
	\centering \includegraphics[width=0.9\linewidth]{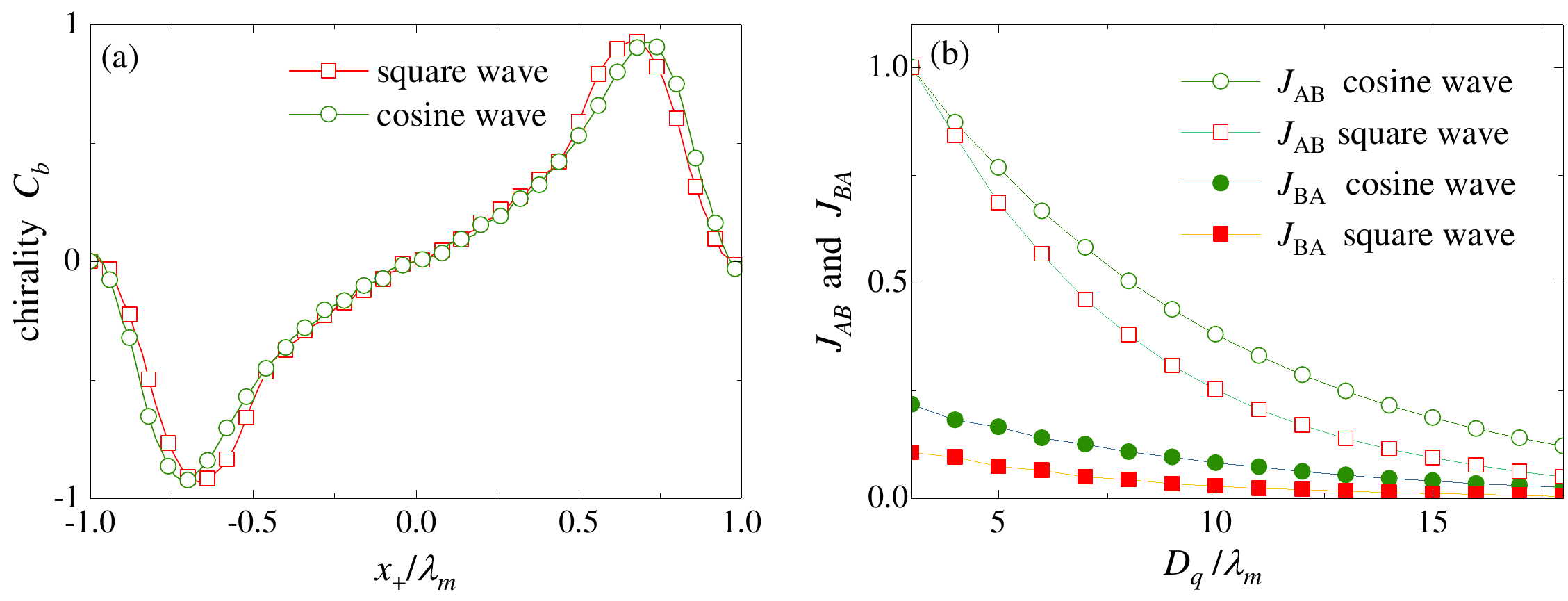}
	\caption{(a) Chirality $\mathcal{C}_b$ as a function of $x_{+}$ for square- and cosine-wave modulations. The parameters used are the same as those in Fig.~3(a) of the main text. (b) Dipole-dipole interaction strengths $J_{AB}$ and $J_{BA}$ as a function of separation $D_q$ for the two modulations. Parameters are adopted from Fig.~5(a) of the main text.
	}
	\label{fig9s}
\end{figure}
%
\subsection{The decay effects of multiple atoms interacting with PCW}
We now show that, with a large detuning $\delta_0$, the atomic decay to the PCW can be effectively suppressed, and the first-order iterative results can well describe the dipole-dipole interactions between atoms.	We start from Eq.~(\ref{Laplace}), which describes the probability amplitudes $C_{\pm}(t)$ ($t>0$) for states $|\Psi_{\pm}\rangle$. It can be exactly derived via the inverse Laplace transform. To go beyond the first-order iterative approximation (i.e., assuming $z\simeq \omega_q+i0^+$), it is convenient to calculate the Laplace transform by using contour integral in the lower half-plane of the complex plane [see Fig.~\ref{fig10s}(b)]. To proceed, we need to calculate the poles of the Green function~\cite{cohen1998atoms,Bello2019s}
\begin{equation}
z - \Sigma_e (z) \mp \Sigma_{AB}(z)=0,
\label{polet}
\end{equation}	
which is a transcendental iterative equation, and cannot be analytically solved. As depicted in Fig.~\ref{fig10s}(a), and explained in previous section, $\Sigma_e (z)$ has the same value for both $G_{\pm}(z)$, which is just the Stark shift $\delta_{qs}$. Therefore, we can simply assume $ \Sigma_e (z)\simeq \delta_{qs}$ as a constant. By replacing $z-\delta_{qs}\rightarrow z$, we have
\begin{equation}
z \mp \Sigma_{AB}(z)=0, \quad \Sigma_{AB}(z)\simeq  \frac{L}{2\pi}\int_{-k_0}^0 d\delta k \frac{2 \Re(g_{kA} g_{kB}^*)}{z +\delta'_0 +\alpha_m \delta k^2},
\label{zpole}
\end{equation}
where $\delta'_0=\delta_0+\delta_{qs}$ is the renormalized detuning. When the coupling $g_{kA}$ and $g_{kB}$ are sufficiently weak, we have $\delta'_0\simeq\delta_0$. Since only the modes around the band edge have significant contributions to the system's dynamics, the couplings can be approximate as $g_{kA}e^{-ikx_A}\simeq g_{kB}e^{-ikx_B}\simeq g_{k0}$. We note that the lower bound of the integral can be extended to $-\delta k_0\simeq -\infty$. With these approximations, $\Sigma_{AB}(z)$ can be written as
\begin{equation}
\Sigma_{AB}(z)\simeq  \frac{\pi g_{0}^{2}}{z+\delta'_{0}}\frac{L} { 2\pi L_{\text{eff}}} e^{-|\frac{D_q}{L_{\text{eff}}}|},   \quad  \sqrt{\frac{\alpha _m}{z+\delta'_{0}}}=L_{\text{eff}}, \quad D_q=x_{A}-x_{B},
\end{equation}
where $2\pi/L=dk$ is the mode discretization space of the PCW. Note that $z \mp \Sigma_{AB}(z)=0$ is still a transcendental equations. The exponential term in $\Sigma_{AB}(z)$ indicates that the qubit-qubit interactions decays as their separation distance $D_q$ increases. This analytical result matches the numerical ones, as shown in Fig.~\ref{fig9s}(b). By taking the first-order iterative results with $z\simeq\omega_q=0$ (note that we work in the rotating frame at the atomic transition frequency), the effective decay length have the same formula in Eq.~(\ref{phib_ana}).

\begin{figure}[tbph]
	\centering \includegraphics[width=0.8\linewidth]{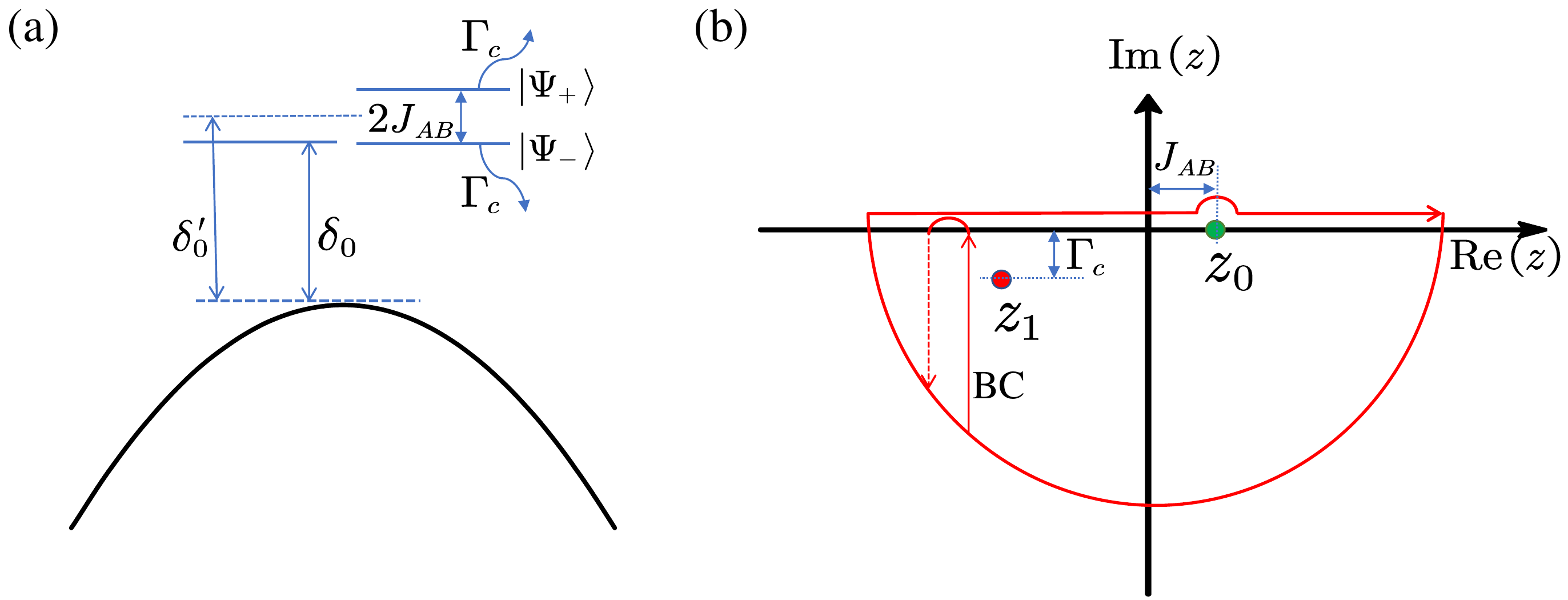}
	\caption{(a) The energy-level diagram for two atoms interacting with the first PCW band. The detuning to the band edge of $|\Psi_{\pm}\rangle$ are slightly renormalized as $\delta'_0$ due to the Stark shift. The splitting between $|\Psi_{+}\rangle$ and $|\Psi_{-}\rangle$ results from the dipole-dipole interaction $J_{AB}$. The decay rate is denoted as $\Gamma_{c}$. (b)  Contour integral used in the calculation $C_{+}(t)$. Both the inside real poles $z_0$ and complex poles $z_1$ contribute to the dynamics. The branch cut (BC) leads to apparent effects when $\delta'_0\simeq 0$, which can be neglected in our discussion.
	}
	\label{fig10s}
\end{figure}

\begin{figure}[tbph]
	\centering \includegraphics[width=0.47\linewidth]{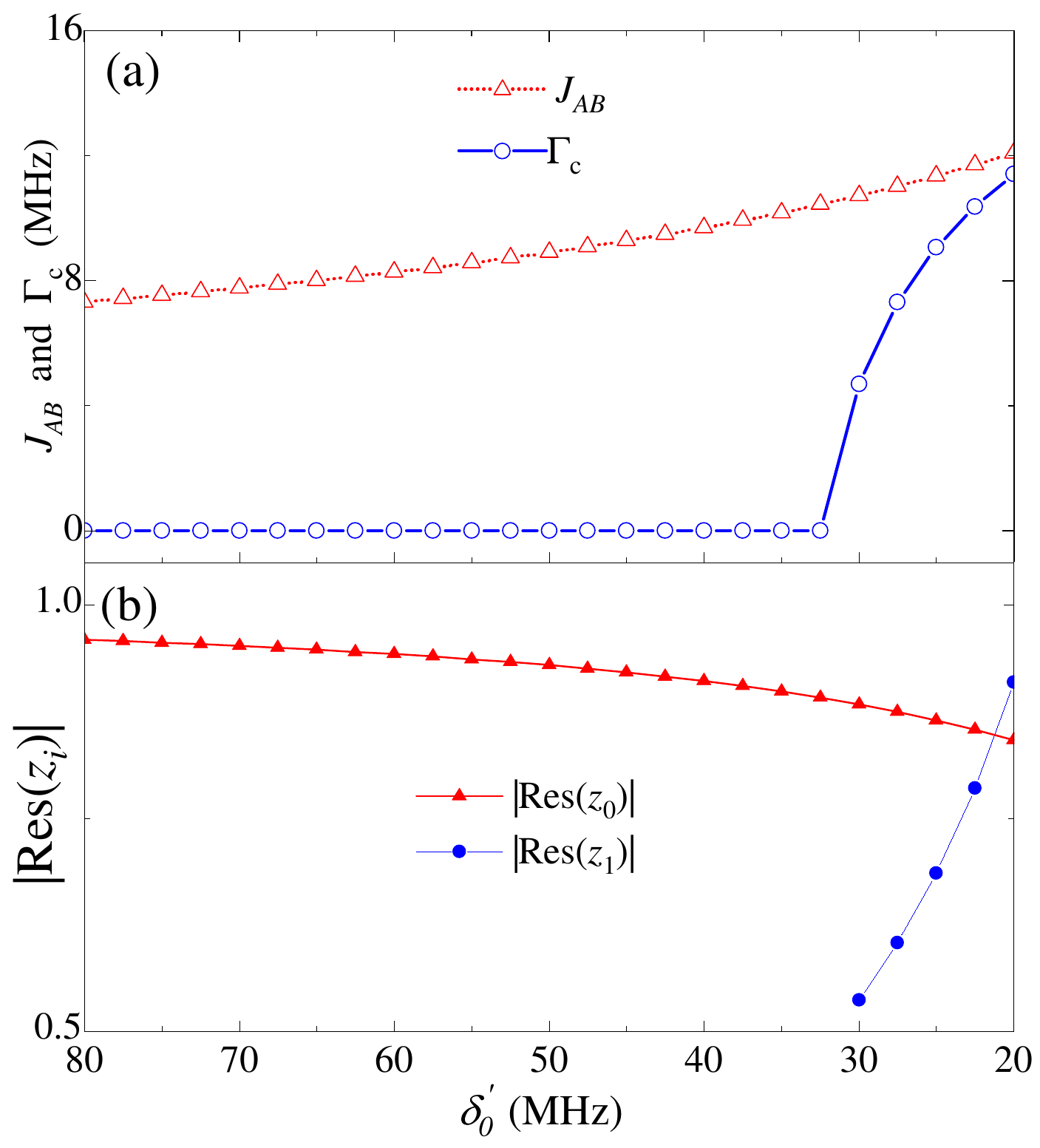}
	\caption{(a) The dipole-dipole coupling strength $J_{AB}$ and the atomic decay rate $\Gamma_c$ versus detuning $\delta_0$. (b) The Residues $|\text{Res}(z_i)|$ of the coherent coupling and decay terms for the evolution $C_{+}(t)$ versus detuning $\delta'_0$.
	}
	\label{fig11s}
\end{figure}

For simplicity, we consider the case where $D_q\ll L_{\text{eff}}$, i.e., the two atoms are close enough compared to the spatical decay length.  $\Sigma_{AB}(z)$ is now simplified as
\begin{equation}
\Sigma_{AB}(z)\simeq \frac{\pi g_{0}^{2}}{\sqrt{\alpha _m\left( z+\delta _{0}^{\prime} \right)}}\frac{L}{2\pi}.	
\end{equation}
As depicted in Fig.~\ref{fig10s}(a), the symmetric Rabi splittings between $|\Psi_{\pm}\rangle$ are due to the coupling between two atoms.
Therefore, we can just take $|\Psi_{+}\rangle$  to calculate $J_{AB}$ and the decay $\Gamma_{c}$. As depicted in Fig.~\ref{fig10s}(b), by solving 
\begin{equation}
z-\Sigma_{AB}(z)=0,
\end{equation}
we obtain two poles for the green funtion $G_{+}(z)$, where $z_{0}$ (on the real axis) and $z_{1}$ (a complex pole with $\text{Im}(z_1)<0$) are both inside the contour for $C_{+}(t)$.  As discussed in Refs.~\cite{GonzlezTudela2017s,Bello2019s,Ramos2016s}, since $\Sigma_{AB}(z)$ is a multivalued function due to the square root term $\sqrt{z+\delta'_0}$, at point $z=-\delta'_0$ one has to introduce branch cuts [BC, red dashed arrows in Fig.~\ref{fig10s}(b)].	This branch cut describes a non-exponential decay process.	However, its contribution only plays an important role when the atom frequency approaches the band edge (i.e., $\delta'_0\simeq 0$)~\cite{GonzlezTudela2017s}. Thus, in our discussions, we can neglect this effect. Consequently, the evolution can be derived via the residue theorem
\begin{equation}
C_{+}(t)\simeq \text{Res}(z_0)e^{-iz_0 t}+\text{Res}(z_1)e^{-iz_1 t},
\end{equation}
where $\text{Res}(z_i)$ are residues of the Green function $G_{+}(z)$
\begin{equation}
\text{Res}(z_i)=\frac{1}{1-\partial_z \Sigma_{AB}(z)}\Big|_{z=z_i}, \quad i=0,1.
\end{equation}
For a small detuning $\delta'_0$, the physical processes are now clear: $|\text{Im}(z_1)|=\Gamma_{c}$ is the atomic decay rate due to coupling to the waveguide, and $z_{0}$ just contributes to a dynamical phase due to the Rabi splitting $J_{AB}\simeq z_0$ between two atoms (without decoherence). The dynamical evolution is described as a fractional decay, where the atomic excitation is partly leaked into the PCW, and the other part is localized~\cite{GonzlezTudela2017s}. The contributions of these two processes are evaluated from their residues $|\text{Res}(z_i)|$. Since the term with $z_1$ is unstable under decay, in the long-time limit, the probability that the atoms remain in the superposition state is~\cite{Bello2019s,Ramos2016s}
\begin{equation}
|C(t\rightarrow \infty)|^2\simeq |\text{Res}(z_0)|^2.
\end{equation}
Given that $|\text{Res}(z_1)|\simeq0$ and $|\text{Res}(z_0)|\simeq1$, the atomic energy leaking into the waveguide approximately vanishes, and the interaction between two atoms is purely coherent. In this case, two atoms remain in the initial superposition state $|\Psi_{+}\rangle$ with $|C(t=\infty)|^2\simeq 1$.

To evaluate $J_{AB}$ and $\Gamma_c$, we adopt the experimentally feasible parameters listed in Table~\ref{table1}. As depicted in Fig.~\ref{fig4s}, the PCW gap is about $\Delta_g/(2\pi)=800~\text{MHz}$. Therefore, we set $\delta'_{0}/(2\pi)\sim0.1\Delta_g=80~\text{MHz}$ in our discussion. Moreover, the coupling is set as $g_{0}= 0.8~\text{MHz}$ with a mode discretization space $dk=10^{-4}k_m$. In Fig.~\ref{fig11s}(a), we plot the coupling strength $J_{AB}\simeq z_{0}$, and the decay rate $\Gamma_{c}\simeq|\text{Im}(z_1)|$ versus $\delta'_{0}$. Their contribution weights $|\text{Res}(z_i)|$ are plotted in Fig.~\ref{fig11s}(b). We thus infer that, for a large detuning, $J_{AB}\gg\Gamma_c$ and $|\text{Res}(z_0)|\simeq 1\gg |\text{Res}(z_1)|$. Note that $|\text{Res}(z_0)+\text{Res}(z_1)|\neq 1$ since we neglect the branch cut contribution~\cite{GonzlezTudela2017s}. After $\delta'_0/(2\pi)>30~\text{MHz}$, the complex pole $z_1$ disappears with $\Gamma_c=0$ and $|\text{Res}(z_0)|\simeq 1$, indicating that the coherent dipole-dipole coupling $J_{AB}$ contributed by $z_0$ dominates the evolution, while the energy leaking to the PCW takes little effect. Additionally, as shown in Fig.~\ref{fig11s}(a), the dipole-dipole coupling is about $J_{AB}/(2\pi)\simeq 8~\text{MHz}$, which is strong enough in circuit-QED for quantum coherent control. Since $J_{AB}\ll \delta'_0$, it is also reasonable to adopt the first-order approximation in Eq.~(\ref{firstoder}), to calculate $J_{AB}$ by neglecting the decay effects.

\subsection{Topological phases with giant atoms}
The chiral dipole-dipole interactions in Fig.~4 of the main text provide an ideal platform to simulate the Su-Schrieffer-Heeger (SSH model), which is described by a one-dimensional Hamiltonian with nontrivial topology~\cite{Su1979s}. The Hamiltonian for the atomic chain reads
\begin{equation}
H_{\text{qc}} = \sum_i (J_{AB} \sigma_{Ai}^- \sigma_{Bi}^+ + J_{BA} \sigma_{Bi}^- \sigma_{Ai+1}^+) + \text{H.c.}, 
\label{dipolec}
\end{equation}
whose bulk spectrum is gapped given that $J_{AB} \neq J_{BA}$~\cite{Ozawa2019s}. The relation between $J_{AB}$ and $J_{BA}$ determines whether the winding number is a nonzero integer or not~\cite{Ryu2010s, Kane2013s, Xiao2014s, SaeiGharehNaz2018s}. The two lowest energy bands of $H_{\text{qc}}$ are characterized by the topological invariant, i.e., the Zak phase $\mathcal{Z}$, and the corresponding relation is~\cite{Ozawa2019s}
\begin{subequations}
	\begin{gather}
	J_{AB} > J_{BA}, \quad \mathcal{Z} = 0, \quad \text{trivial insulator}, \\
	J_{AB} < J_{BA}, \quad \mathcal{Z} = \pi, \quad \text{nontrivial insulator}, \label{nontopo}
	\end{gather}
\end{subequations}
where the critical point $J_{AB} = J_{BA}$ corresponds to the topological phase-transition point~\cite{SaeiGharehNaz2018s}.
In the topologically nontrivial phase with $J_{AB} < J_{BA}$, there are zero-energy edge modes located at two ends of the finite chain, whose energy spectra are isolated and topologically protected from the bulk modes. In the topologically trivial phase with $J_{AB} > J_{BA}$, such edge modes do not exist.
In experiments, the topological invariant is identified by the topological phase-transition process~\cite{Lohse2015s, Gu2017arxivs, Nakajima2016s, SaeiGharehNaz2018s}.
Realizing the transition between the topologically trivial and nontrivial phase of the SSH model requires tuning all coupling strengths simultaneously, as well as reversing the relation between $J_{AB}$ and $J_{BA}$, which is very challenging in experiments~\cite{Xiao2014s, SaeiGharehNaz2018s}. 

As shown in Fig.~5(b) of the main text, such a topological transition can be easily realized by shifting the modulation signal of the PCW with a distance $d_s$. 
The impedance of the Josephson PCW is modulated via external flux signals instead of being fabricated with unchangeable parameters. Shifting the PCW modulation signal will change the interference relations and the bound-state chirality.
As depicted in Fig.~4(a,b) of the main text, by shifting the programmable modulating signal a certain distance $d_s$, the highest-impedance positions will also be moved.
The phase transition point is at $d_s = 0.25\lambda_m$, around which $J_{AB}$ ($J_{BA}$) decreases (increases) linearly with $d_s$. By changing the flux $\Phi_{\text{q}}$ through each atom's split loop, the qubit frequency can also be modulated in time~\cite{Gu2017s}. 

We can map the SSH chain to the tight-binding Rice-Mele (RM) model~\cite{Lohse2015s, Gu2017arxivs, Nakajima2016s}:
\begin{equation}
H_{\text{qc}} = \sum_i \mleft[J_{AB}(t) \sigma_{Ai}^- \sigma_{Bi}^+ + J_{BA}(t) \sigma_{Bi}^- \sigma_{Ai+1}^+ + \text{h.c.} \mright] + \sum_i \Delta_q (t) (\sigma_{Ai}^z - \sigma_{Bi}^z). 
\end{equation}
In Fig.~6(a) of the main text, the degenerate point of the RM model is at $\{J_{BA} - J_{AB}, \Delta_{q}\} = \{0, 0\}$, which is also the phase-transition point of the SSH model. As discussed in Refs.~\cite{Lohse2015s, Gu2017arxivs, Nakajima2016s}, all the adiabatic quantum pump trajectories which encircle the degeneracy point are topologically equivalent, and robust to disorder and perturbations.

In our proposal, the coupling difference $J_{AB} - J_{BA}$ linearly depends on the signal shifting the distance $d_s$. In experiments, one can adiabatically modulate $d_s$ back and forth in cosine form. Moreover, the qubit frequencies can be tuned in the sine form. Therefore, we can assume
\begin{equation}
J_{AB}(t) = 1 - \delta_{\alpha}\cos\left(\frac{2\pi t}{T}\right), \quad 
J_{BA}(t) = 1 + \delta_{\alpha}\cos\left(\frac{2\pi t}{T}\right), \quad \Delta_q(t) = \Omega_p \sin\left(\frac{2\pi t}{T}\right).
\label{loop}
\end{equation}

As depicted in Fig.~3 in the main text, the maximum chirality of the bound state is about $0.95$, indicating that $J_{AB}$ or $J_{BA}$ cannot be exactly zero. However, the topological pumping processes encircling the degeneracy point
$\{J_{BA} - J_{AB}, \Delta_{q}\}=\{0, 0\}$ are topologically equivalent [see Fig.~6(b) in the main text], and robust to disorder and perturbations. Therefore, we just require $J_{BA} - J_{AB}$ (rather than $J_{AB}$ and $J_{BA}$) to vary across zero.
In Fig.~6(b) of the main text, by assuming an SSH chain with site number $N=12$, and setting the parameters as: $\delta_{\alpha} = 0.9$, $\Omega_p = 0.3$, and $T = 100$, we plot the evolution of an initial excitation localized at the first site on the left edge. The minimum values for $J_{AB}$ and $J_{BA}$ are nonzero and equal to $0.1$. As depicted in Fig.~6(b), the excitation is transferred to the right edge state at the end of each pump circle (without being disturbed by a nonzero coupling strength), and this adiabatic process is topologically protected.

\end{document}